\documentclass[%
 reprint,
superscriptaddress,
 amsmath,amssymb,
 aps,
 prl,
]{revtex4-2}

\usepackage{graphicx}
\usepackage{todonotes}
\usepackage{float}
\usepackage[version=4]{mhchem}
\usepackage{dcolumn}
\usepackage{bm}
\usepackage{hyperref}
\hypersetup{
   breaklinks=true,   
   colorlinks=true,   
}

\begin{document}
\preprint{APS/123-QED}
\title{Exploring Isospin Symmetry Breaking in Exotic Nuclei: High-Precision Mass Measurement of $^{23}$Si and Shell-Model Calculations of $T=5/2$ Nuclei}


\author{F.~M.~Maier}
\email{maierf@frib.msu.edu}
\affiliation{Facility for Rare Isotope Beams, East Lansing, Michigan, 48824, USA}
\author{G.~Bollen}
\affiliation{Facility for Rare Isotope Beams, East Lansing, Michigan, 48824, USA}
\affiliation{Department of Physics and Astronomy, Michigan State University, East Lansing, Michigan 48824, USA}
\author{B.~A.~Brown}
\affiliation{Facility for Rare Isotope Beams, East Lansing, Michigan, 48824, USA}
\affiliation{Department of Physics and Astronomy, Michigan State University, East Lansing, Michigan 48824, USA}
\author{S.~E.~Campbell}%
\affiliation{Facility for Rare Isotope Beams, East Lansing, Michigan, 48824, USA}
\affiliation{Department of Physics and Astronomy, Michigan State University, East Lansing, Michigan 48824, USA}
\author{X.~Chen}
\affiliation{Facility for Rare Isotope Beams, East Lansing, Michigan, 48824, USA}
\author{H.~Erington}
\affiliation{Facility for Rare Isotope Beams, East Lansing, Michigan, 48824, USA}
\affiliation{Department of Physics and Astronomy, Michigan State University, East Lansing, Michigan 48824, USA}
\author{N.~D.~Gamage}
\affiliation{Facility for Rare Isotope Beams, East Lansing, Michigan, 48824, USA}
\author{C.~M.~Ireland}
\affiliation{Facility for Rare Isotope Beams, East Lansing, Michigan, 48824, USA}
\affiliation{Department of Physics and Astronomy, Michigan State University, East Lansing, Michigan 48824, USA}
\author{R.~Ringle}
\affiliation{Facility for Rare Isotope Beams, East Lansing, Michigan, 48824, USA}
\affiliation{Department of Physics and Astronomy, Michigan State University, East Lansing, Michigan 48824, USA}
\author{S. Schwarz}
\affiliation{Facility for Rare Isotope Beams, East Lansing, Michigan, 48824, USA}
\author{C.~S.~Sumithrarachchi}
\affiliation{Facility for Rare Isotope Beams, East Lansing, Michigan, 48824, USA}
\author{A.~C.~C.~Villari}
\affiliation{Facility for Rare Isotope Beams, East Lansing, Michigan, 48824, USA}
\date{\today}%

\begin{abstract}
\noindent
We present a high-precision mass measurement of the proton-rich nucleus $^{23}$Si, performed with the LEBIT Penning trap at the Facility for Rare Isotope Beams (FRIB) utilizing the time-of-flight ion cyclotron resonance (TOF-ICR) technique. We determined a mass excess of 23362.9(5.8)~keV, which agrees with a recent storage-ring measurement from CSRe but has a factor of 20 improved precision. $^{23}$Si is hence the nucleus with the most precisely known mass of all nuclei with an isospin projection of $T_z = -5/2$. We performed shell-model calculations with the USDC and USDCm Hamiltonians to study binding energy differences and Thomas-Ehrmann shifts in mirror systems with an isospin up to $T=5/2$. Our experimental result and other recently reported masses of neutron-deficient sd-shell nuclei agree well with the theoretical predictions, demonstrating that isospin symmetry breaking in sd-shell nuclei — even at high isospin values — is well described by modern shell-model calculations.
\end{abstract}

\maketitle

\section{I. Introduction}
\begin{figure}[t]
\includegraphics[width=\columnwidth]{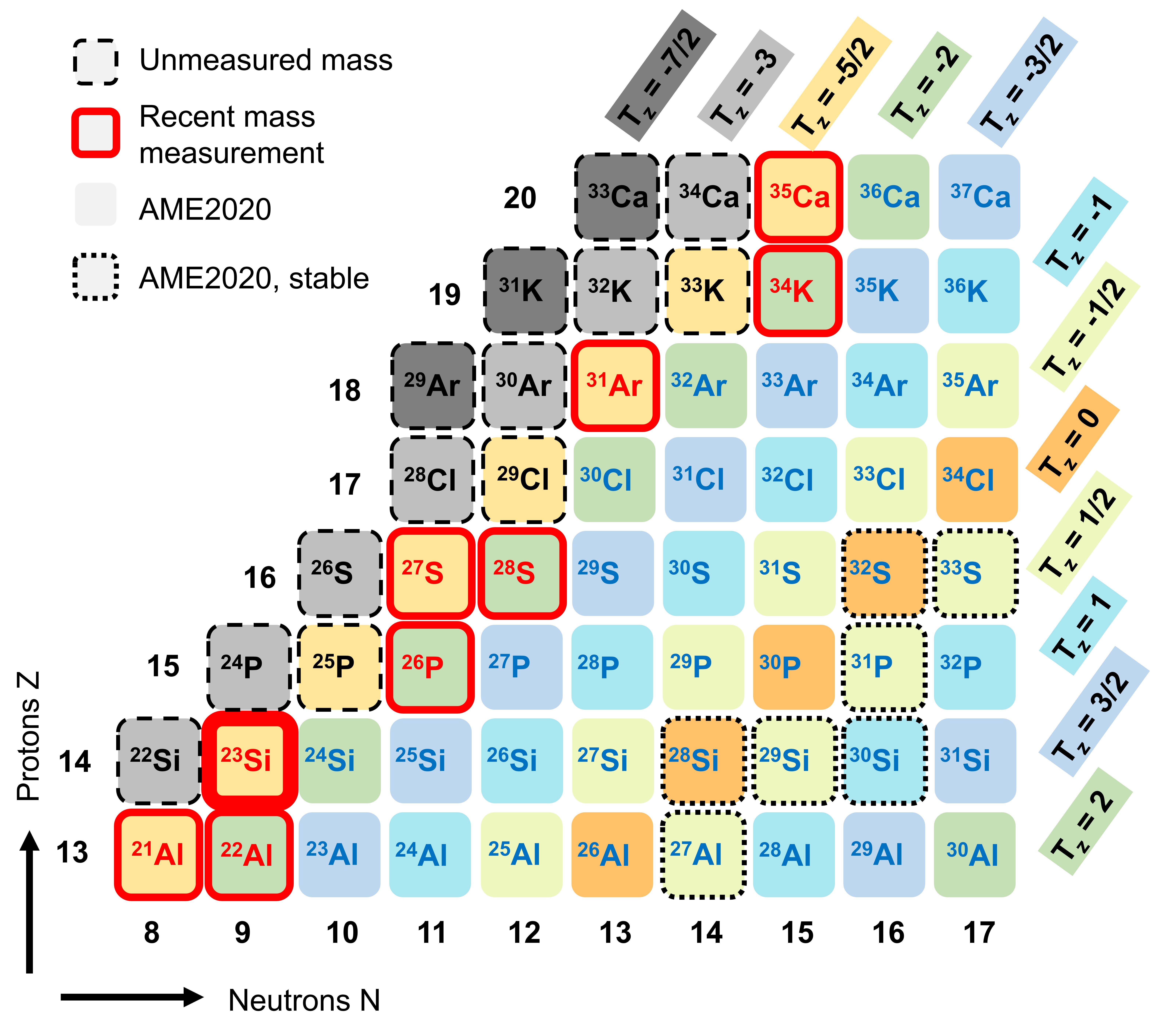}
\caption{The nuclear landscape in the region of $^{23}$Si. Individual nuclei are colored based on their respective isospin~$T$. A black dashed border refers to an unmeasured mass, a black dotted border to a stable nucleus, no border to a mass measurement reported in AME2020~\cite{AME2020} and a red full border to a recent mass measurement as reported in this work for $^{23}$Si and references~\cite{PhysRevC.110.L031301, PhysRevLett.133.222501,
PhysRevLett.131.092501, PhysRevLett.132.152501,PhysRevC.110.L031302}.} 
\label{fig:nuclearchart}
\end{figure}

Atomic nuclei are fascinating quantum many-body systems composed of two types of fermions: protons and neutrons. Protons and neutrons have nearly identical masses and demonstrate similar behaviors when interacting through the strong nuclear force. In the 1930s, Heisenberg~\cite{Heisenberg1932} and Wigner~\cite{PhysRev.51.106} introduced the concept of isospin, which revolutionized the understanding of nuclear forces and has become a cornerstone in the theoretical modeling of atomic nuclei.
The isospin quantum number $T$ describes the symmetry of nuclear states under the strong force. Both protons and neutrons are assigned the same isospin value of $T=1/2$, but differ in their isospin projection: protons have $T_z=-1/2$, while neutrons have $T_z=1/2$. For a given nucleus with total isospin $T$, the isospin projection is given by $T_z=(N-Z)/2$, where $N$ is the number of neutrons and $Z$ is the number of protons. Fig.~\ref{fig:nuclearchart} shows the proton-rich side of the nuclear chart for $Z=13$ to $Z=20$ with nuclei having the same isospin $T$ depicted in the same color. Pairs of nuclei with the same total number of nucleons but exchanged proton and neutron number, e.g. $^{30}$Cl and $^{30}$Al, are called mirror nuclei. Assuming perfect isospin symmetry, the difference between the binding energies of mirror partners is zero. 
However, while isospin symmetry provides a useful approximation, it is not exact. One of the primary sources for this symmetry breaking is the Coulomb interaction between the protons~\cite{physics5020026, sym16060745} causing a small, but measurable, difference in the binding energy of mirror nuclei. Furthermore, other effects, such as those arising from charge-dependent nucleon-nucleon interactions~\cite{physics5020026,sym16060745}, contribute to further symmetry breaking. 

Taking these symmetry breaking effects into account, the concept of isospin remains a powerful tool for predicting excitation energies and ground-state masses of exotic nuclei that are challenging to probe experimentally. This predictive capability is exemplified by the isobaric multiplet mass equation (IMME), which relates the mass excess $M$ of the $2T+1$ nuclear states of a given isobaric multiplet according to
\begin{equation}
    M(\alpha,T,T_z)=a(\alpha,T)+b(\alpha,T)T_z+c(\alpha,T)T_z^2, \label{eq: IMME}
\end{equation}
where $a$, $b$ and $c$ are fitting parameters and $\alpha$ describes the spin-parity of the state. The $2T+1$ nuclear states of a given isobaric multiplet are also often referred to as isobaric analog states. They have the same isospin $T$, the same mass number $A$ and the same spin-parity $J^\pi$ but a different isospin projection $T_z$. 
The IMME is used to predict masses for modeling stellar evolution~\cite{PhysRevC.79.045808,PhysRevC.87.065803,PhysRevC.95.055806}, as well as to estimate the location of the proton dripline, which is essential to understand the limits of nuclear stability~\cite{sym16060745}. Moreover, isospin symmetry and its breaking also play a significant role in particle physics, especially in testing the unitarity of the Cabibbo-Kobayashi-Maskawa (CKM) matrix~\cite{CIRIGLIANO2023137748, PhysRevC.78.035501, PhysRevC.80.064319,SENG2023137654}. In this way, the study of isospin symmetry breaking is not only relevant for nuclear physics but also deepens our understanding of particle physics and the intricate relationships between the forces that govern both the atomic nucleus and subatomic particles.

In recent years, advances in experimental techniques have provided new data on exotic nuclei far from stability with high isospin values, see Fig.~\ref{fig:nuclearchart}. In particular, the masses of several $T_z=-5/2$ nuclei have only very recently been published, such as $^{35}$Ca~\cite{PhysRevLett.131.092501} in 2023 and $^{21}$Al~\cite{PhysRevC.110.L031301}, $^{23}$Si~\cite{PhysRevLett.133.222501},  $^{27}$S~\cite{PhysRevLett.133.222501} and $^{31}$Ar~\cite{PhysRevLett.133.222501} in 2024. These new masses, along with the high-precision mass measurement of $^{23}$Si presented in this work, provide a rich data set to explore isospin symmetry breaking in nuclei with high isospin. The mass of $^{23}$Si was measured at the Low Energy Beam and Ion Trap (LEBIT) at the Facility for Rare Isotope Beams (FRIB) and improves the previous precision by a factor of 20. 
By comparing the experimental binding energy differences of mirror nuclei up to $T=5/2$ with shell-model calculations~\cite{PhysRevC.101.064312}, we can refine our understanding of the Coulomb force's role and other isospin symmetry breaking effects and contribute to the broader effort to decode the intricate dynamics of atomic nuclei.

\section{II. Experimental Method and Analysis}
The ions of interest were produced at FRIB by projectile fragmentation. After acceleration to an energy of 290 MeV/u in FRIB's superconducting Linear Accelerator~\cite{York-LINAC} the primary $^{28}$Si$^{14+}$ beam impinged upon a 17.67~mm thick $^{12}$C target creating a wide variety of different ion species including $^{23}$Si$^{14+}$. The ions of interest were separated by the Advanced Rare Isotope Separator (ARIS)~\cite{Hausmann-ARIS} from most of the other reaction products, as well as the primary beam, and sent to the Advanced Cryogenic Gas Stopper (ACGS)~\cite{Lund-ACGS}. The momentum of the purified beam was compressed using a 1004~$\mu$m thick Al wedge at an opening angle of 5~mrad as well as a 7899~$\mu$m thick Al degrader at an angle of 41~degrees in the last energy-dispersive beamline section leading to ACGS. These settings were optimized for another experiment, in which fully-stripped $^{23}$Si was present as an isotone. Afterwards the beam was stopped in the gas stopper via collisions with helium buffer gas. To guide the ions to the extraction orifice, radio-frequency (RF) carpet surfing~\cite{BOLLEN-IonSurfing} was used. After extraction, the ions were passed through a radiofrequency quadrupole acting as a beam cooler and differential pumping barrier, accelerated to a beam energy of 30 keV and mass-separated by a dipole magnet with a mass resolving power of $\approx 1500$. A continuous beam consisting only of $^{23}$Si$^+$, $^{23}$Al$^+$, $^{23}$Mg$^+$ and $^{23}$Na$^+$ was then sent towards LEBIT. At LEBIT, the ions were injected into a linear Paul-trap cooler-buncher~\cite{CoolerBuncher} which contained helium buffer gas to facilitate accumulation, cooling and bunching of the ions. After a cooling time of $\approx 15$~ms, the ions were extracted from the Paul trap as well-defined ion bunches with significantly reduced longitudinal and transversal emittance. Subsequently, they were guided into LEBIT's 9.4~T hyperbolic Penning trap for further manipulation and analysis~\cite{PenningTrap}. In the Penning trap, the ions were confined in three dimensions by a superposition of a magnetic field $B$ and an electrostatic quadrupole field. The motion of the ion is described by three distinct eigenfrequencies: an axial frequency $\nu_{z}$ and two radial frequencies, $\nu_{-}$ (magnetron frequency) and $\nu_{+}$ (reduced cyclotron frequency). In ideal conditions, the cyclotron frequency $\nu_{c}$ is given by the sum of the two radial frequencies, $\nu_{c}=\nu_{+} + \nu_{-}$~\cite{Gabrielse-Sideband}. It is related to the ion's mass-over-charge ratio $m/q$ via 
\begin{equation}\label{eq:nuc}
  \nu_{c} = \frac{q}{m} \frac{B}{2\pi}.  
\end{equation}
To account for uncertainties and drifts in the magnetic field, measurements of $\nu_{c}$ of $^{23}$Si$^+$ were interleaved with those of the well-known stable reference $^{23}$Na$^+$, $\nu_{c, \mathrm{ref}}$. The cyclotron frequency ratio $R$ is then given by 
\begin{equation}\label{eq:mass_from_ratio}
   R = \frac{ \nu_{c} }{ \nu_{c,\textrm{ref}} }.
\end{equation}
The mass of $^{23}$Si was calculated by combining equations \ref{eq:nuc} and \ref{eq:mass_from_ratio} taking  the mass $m_{\mathrm{ref}}$ of the reference ion $^{23}$Na into account and accounting for the mass of the missing electron $m_e$, 
\begin{equation}\label{eq:mass final}
   m = \frac{1}{R} (m_{\mathrm{ref}}-m_e) + m_e.
\end{equation}

\begin{figure}[t]
\includegraphics[width=\columnwidth]{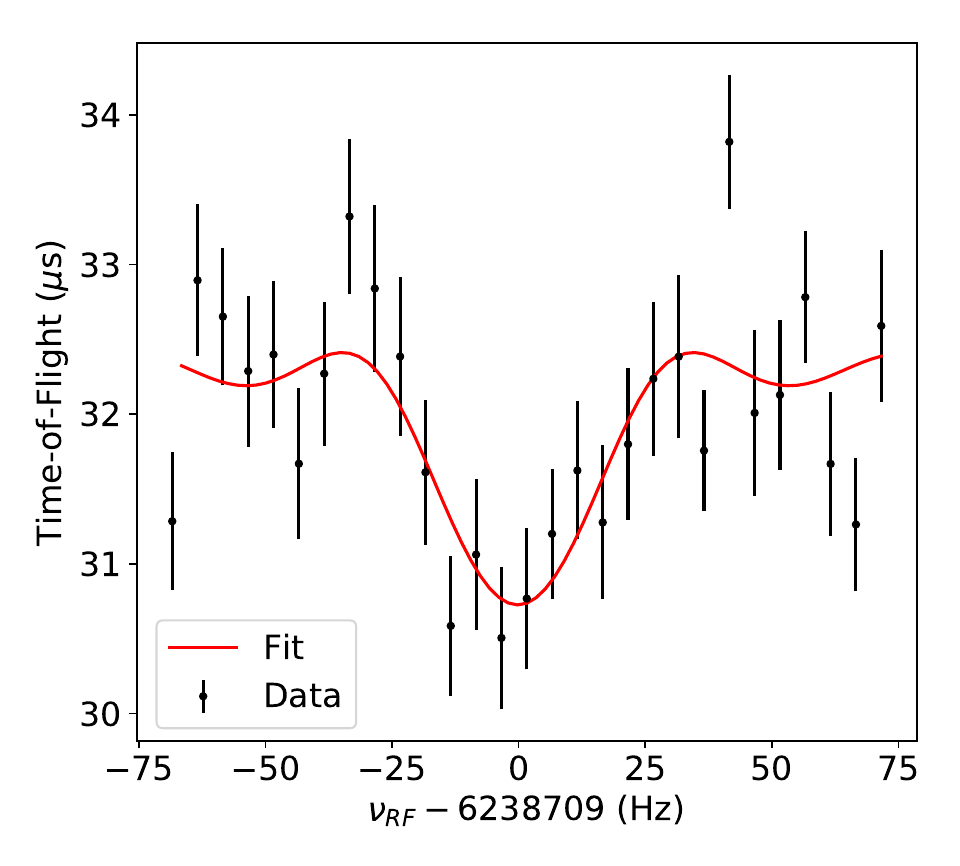}
\caption{A summed ToF-ICR spectrum of the last two cyclotron frequency measurements (number 4 and 5) taken for \textsuperscript{23}Si\textsuperscript{+} with a quadrupole excitation time of 25~ms. The spectrum is formed by 1598 ions. The full red line shows a $\chi^{2}$-minimization fit to the data points depicted in black as described in \cite{KONIG199595}. The cyclotron frequency $\nu_{c}$ is obtained from $\nu_{RF}$ at the minimum time-of-flight of the fit. }
\label{fig:23Si_ToFICR_summed}
\end{figure}

In this study, the time-of-flight ion cyclotron resonance (ToF-ICR) technique~\cite{ToF1,BECKER199053,KONIG199595} was used to measure $\nu_c$. After extraction from the Paul-trap cooler-buncher, the ions were directed off center by a Lorentz steerer~\cite{LorentzSteerer} to produce an initial magnetron radius once the ions are trapped in the Penning trap. In the Penning trap, the ion sample was first cleaned against the remaining isobaric contamination by employing dipolar RF excitation for 15 ms near the respective frequencies of the contaminating species. Afterward, a quadrupole RF excitation pulse was applied with a frequency near the expected cyclotron frequency of the respective nucleus. An excitation time of 250~ms was chosen for the stable reference $^{23}$Na. Considering $^{23}$Si only has a half-life of 42.3 ms~\cite{SHAMSUZZOHABASUNIA20211}, the quadrupole excitation time for $^{23}$Si was reduced to 5 or 25~ms.  The quadrupole excitation pulse converted the slow magnetron motion into a fast reduced cyclotron motion. The ions were then ejected from the Penning trap toward a microchannel plate (MCP) detector located further downstream. The frequency of the quadrupole pulse $\nu_{\mathrm{RF}}$ was scanned resulting in a variable conversion of the magnetron motion to cyclotron motion and the ions' time-of-flight to the MCP detector was recorded. The relationship between time-of-flight and the applied quadrupolar frequency $\nu_{\mathrm{RF}}$ is shown in Fig.~\ref{fig:23Si_ToFICR_summed}. As $\nu_{\mathrm{RF}}$ approached the cyclotron frequency $\nu_c$, the radial energy of the ions increased, resulting in a shorter flight time. The minimum flight time occurred when $\nu_{\mathrm{RF}}$ matched $\nu_c$, enabling a determination of $\nu_c$.

Five cyclotron frequency measurements of $^{23}$Si$^+$ were interleaved with measurements of the stable reference $^{23}$Na$^+$ within a total measurement time of 5 hours. For the first measurement of $^{23}$Si$^+$, the quadrupole excitation time was 5 ms in order to quickly provide an identification of $^{23}$Si, for the remaining measurements 25~ms excitation time was chosen to improve the precision. The first measurement of $^{23}$Si$^+$ lasted 15 minutes, while the remaining measurements lasted 1 hour each. On average 0.2 $^{23}$Si$^+$ ions/s were detected on the MCP detector, for the first measurement we hence had 119 detected ions and for the remaining four measurements 720 detected ions on average. Each individual measurement of the stable reference $^{23}$Na$^+$ took approximately 3 minutes. The number of injected $^{23}$Na$^+$ ions was capped at 2 ions/s to minimize Coulomb interaction of the ions stored in the Penning trap~\cite{ToF1}.

\begin{figure}[t]
\includegraphics[width=\columnwidth]{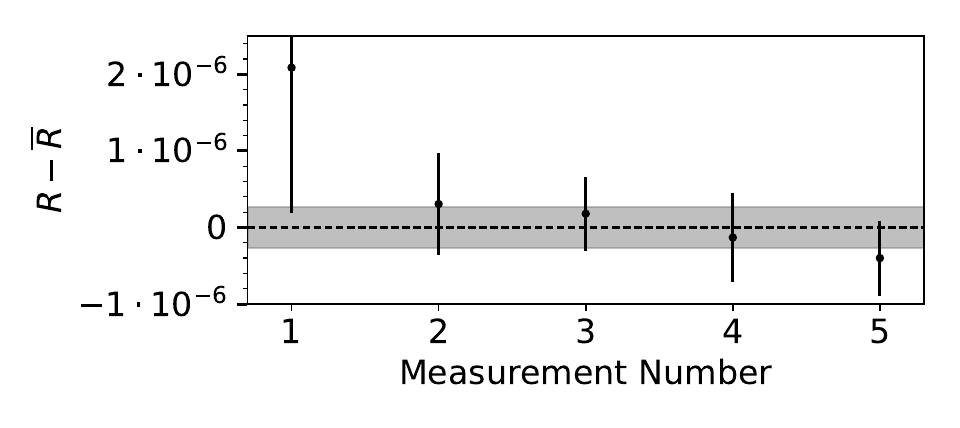}
\caption{Cyclotron frequency ratios $R$ with respect to the average ratio $\bar{R} = 0.99846634(27)$. The gray bar shows the $\pm 1 \sigma$ uncertainty in $\bar{R}$.}
\label{fig:Ratios_23Si}
\end{figure}

\section{III. Results}
The measured frequency ratios $R$ of $^{23}$Si$^+$ and $^{23}$Na$^+$ relative to the weighted average cyclotron frequency ratio $\bar{R}=0.99846634(27)$ are shown in Fig.~\ref{fig:Ratios_23Si}. 
The mass excess of $^{23}$Si follows as 23362.9(5.8)~keV, see Eq.~\ref{eq:mass final}. Systematic errors have been studied in great detail in previous work and are negligible compared to the statistical error: Mass-dependent shifts of the cyclotron frequency caused by magnetic field inhomogeneity and imperfections in the trap lead to an uncertainty of approximately $\delta_{R} \approx 2 \times 10^{-10}/u$~\cite{MassOffsetError}. Nonlinear time-dependent shifts in the magnetic field contribute to an additional uncertainty of $\delta_{R} < 10^{-9}$ per hour \cite{MagneticFieldShift}. To mitigate these, regular reference measurements were conducted. The cyclotron frequency ratio $R$ was regularly compared to the expected $R$ of possible (molecular) isobars. No such isobars were present within the uncertainty limit, thus validating that the measured ions were $^{23}$Si$^+$. 

\begin{figure}[t]
\includegraphics[width=\columnwidth]{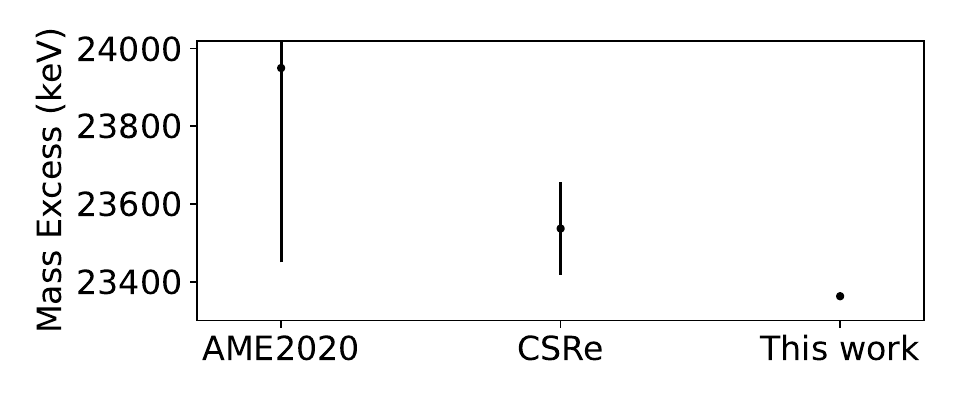}
\caption{Mass excess for $^{23}$Si compared with the extrapolated value from AME2020~\cite{AME2020} and the recent measurement by CSRe~\cite{PhysRevLett.133.222501}.}
\label{fig:Comp_our_prev}
\end{figure}

Our measured mass excess value of $^{23}$Si, 23362.9(5.8)~keV, represents a factor 20 improvement in precision compared to the previous measurement performed at the Cooler-Storage Ring (CSRe) in Lanzhou, which was based on only seven total counts~\cite{PhysRevLett.133.222501}. Our measurement shows that $^{23}$Si is 174(120) keV more bound than the CSRe and 587(500) keV more bound than the extrapolation in AME2020~\cite{AME2020} (see Fig.~\ref{fig:Comp_our_prev} and Tab.~\ref{tab: MassExcessTable}). 

\begin{table}[H] 
\centering
\caption{Mass excess for $^{23}$Si compared with the extrapolated value from AME2020~\cite{AME2020} and the recent measurement by CSRe~\cite{PhysRevLett.133.222501}.}
\vspace{\baselineskip}
\renewcommand{\arraystretch}{1.25}
\setlength{\tabcolsep}{12pt}
\begin{tabular}{l c }
\hline
 & $M_{E}$ (keV)  \\ \hline
AME2020 (extrapolated)   & $23950(500)$       \\ 
CSRe    & $ 23537(119)  $        \\ 
This work    &$23362.9(5.8) $      \\ 
\hline
\end{tabular}
\label{tab: MassExcessTable}
\end{table}

\section{IV. Discussion}
\begin{figure}[t]
\includegraphics[width=\columnwidth]{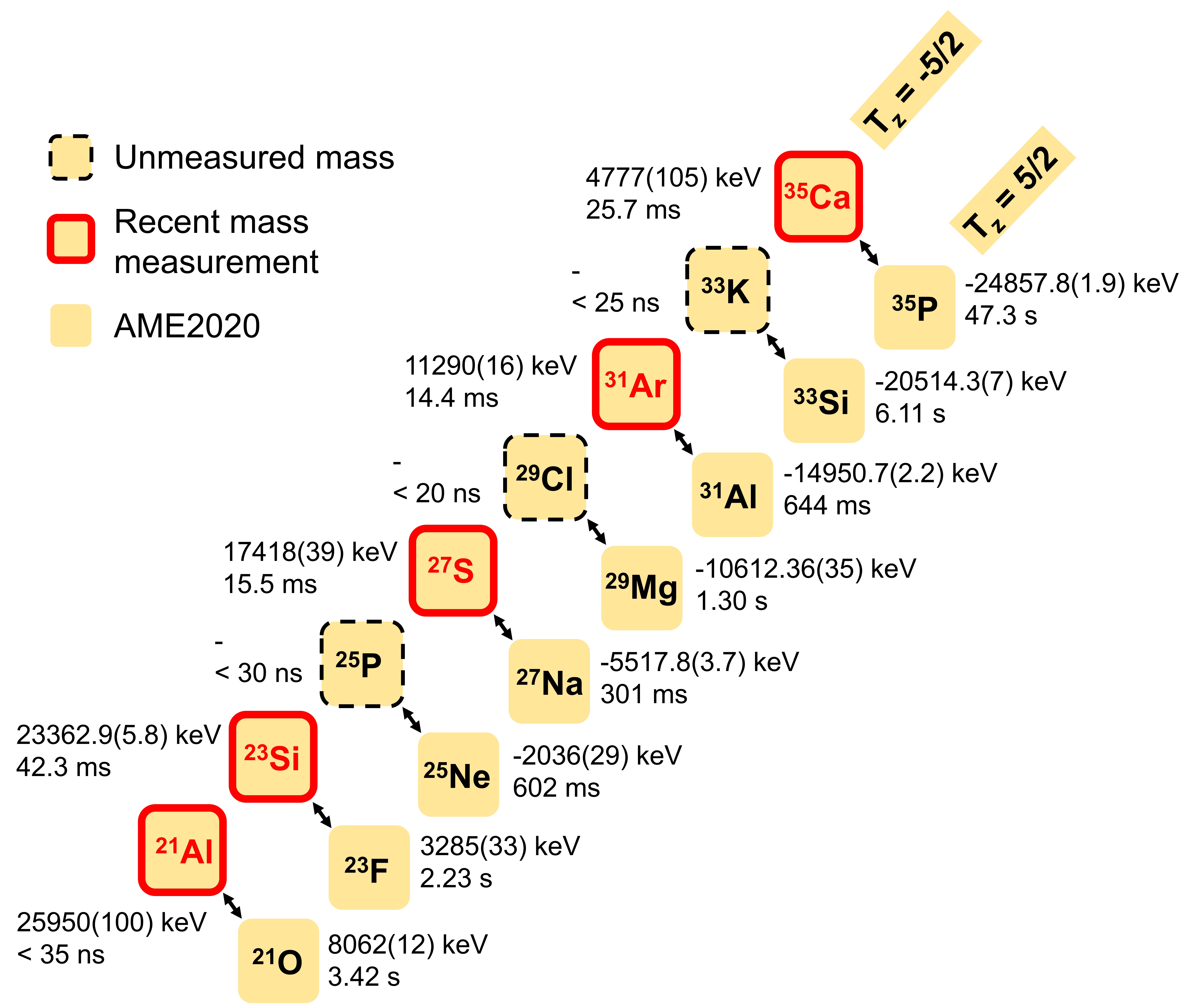}
\caption{Nuclei with isospin $T=5/2$ in the $sd$ shell: The mass excess values are taken from AME2020~\cite{AME2020} except for $^{21}$Al~\cite{PhysRevC.110.L031301}, $^{23}$Si (this work), $^{27}$S~\cite{PhysRevLett.133.222501}, $^{31}$Ar~\cite{PhysRevLett.133.222501} and $^{35}$Ca~\cite{
PhysRevLett.131.092501}. A black dashed border around a nuclide refers to an unmeasured mass, a full red border to a recent mass measurement and no border to a mass measurement as reported in AME2020~\cite{AME2020}.}
\label{fig:isotopesTis5half}
\end{figure}

Thanks to our new measurement, $^{23}$Si is the nucleus with the most precisely known mass of all the $T_z=-5/2$ nuclei. The difference in binding energy between $^{23}$Si and its mirror partner $^{23}$F is given by subtracting the ground state binding energy of $^{23}$Si (151324.5(5.8)~keV) from the ground state binding energy of $^{23}$F (175314(34)~keV)~\cite{AME2020} amounting to $\Delta E = 23.984(31)$~MeV. Together with other recent mass measurements of $T_z=-5/2$ and $T_z=-2$ nuclei, see Figs.~\ref{fig:nuclearchart} and \ref{fig:isotopesTis5half}, and the mass measurements reported in AME2020~\cite{AME2020}, a rich data set of binding energies was compiled and compared with shell-model calculations employing the USDC Hamiltonian~\cite{PhysRevC.101.064312}. This Hamiltonian was derived for $sd$ shell nuclei that explicitly contained isospin-breaking interactions. It was assumed that a single `universal' set of single-particle energies (SPE) and two-body matrix elements (TBME) can be used for all nuclei with $Z$ and $N$ between 8 and 20. The strong interaction TBME were taken to have a smooth mass dependence proportional to $(18/A)^{0.3}$. The TBME for the Coulomb interaction were calculated with harmonic-oscillator radial wavefunctions for $^{28}$Si with $\hbar\omega =12.1$~MeV. These were scaled for other nuclei with a smooth mass dependence of the form $(28/A)^p$ with $p=1/6$ as expected for a potential with $V(r)=1/r$ evaluated with $\hbar \omega$ proportional to $A^{-1/3}$. The Coulomb interaction could be isolated by the energy differences of the mirror nuclei, $\Delta E$. These are connected to the $b$-coefficient of the IMME, see equation~\ref{eq: IMME}, by $b=\Delta E/(2T)$. With the fixed Coulomb TBME, the three SPE associated for the Coulomb interaction were adjusted to fit the $b$-coefficients with a resulting rms difference between experiment and theory of 65 keV (see Fig. 6 in Ref.~\cite{PhysRevC.101.064312}). 

\begin{figure}[t]
\includegraphics[width=\columnwidth]{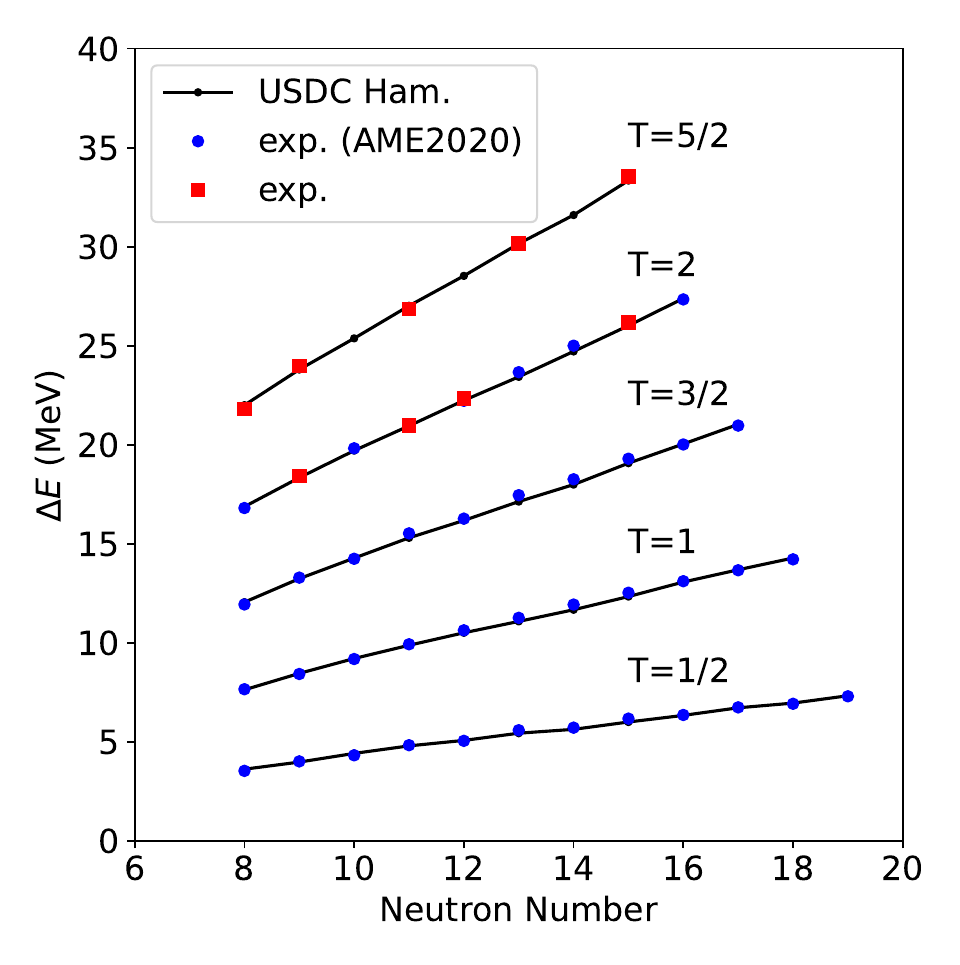}
\caption{Binding energy differences $\Delta E$ as a function of the neutron number of the mirror pair nucleus with $T_z=-T$: the experimental data points shown as blue dots are calculated from the mass measurements reported in AME2020 and the experimental data points shown as red squares are calculated based on AME2020 as well as the recent mass measurements of $^{21}$Al~\cite{PhysRevC.110.L031301}, $^{23}$Si (this work), $^{27}$S~\cite{PhysRevLett.133.222501}, $^{31}$Ar~\cite{PhysRevLett.133.222501} and $^{35}$Ca~\cite{
PhysRevLett.131.092501} for $T=5/2$ as well as $^{22}$Al~\cite{PhysRevLett.132.152501}, $^{26}$P~\cite{PhysRevLett.133.222501}, $^{28}$S~\cite{PhysRevLett.133.222501} and $^{34}$K~\cite{PhysRevC.110.L031302} for $T=2$. The errors are smaller than the dots. In black corresponding binding energy differences from shell-model calculations employing the USDC Hamiltonian~\cite{PhysRevC.101.064312} are shown. }
\label{fig:Mirrorenergydifferences}
\end{figure}

The experimental and theoretical binding energy differences $\Delta E$ for the $sd$-shell ground states are shown in Fig.~\ref{fig:Mirrorenergydifferences} for the respective mirror pairs from $T=1/2$ to $T=5/2$. The present result for the mirror pair $^{23}$Si - $^{23}$F and other experimental results obtained after the USDC Hamiltonian was established in 2020 are depicted as red squares. 

\begin{figure}[t]
\includegraphics[width=\columnwidth]{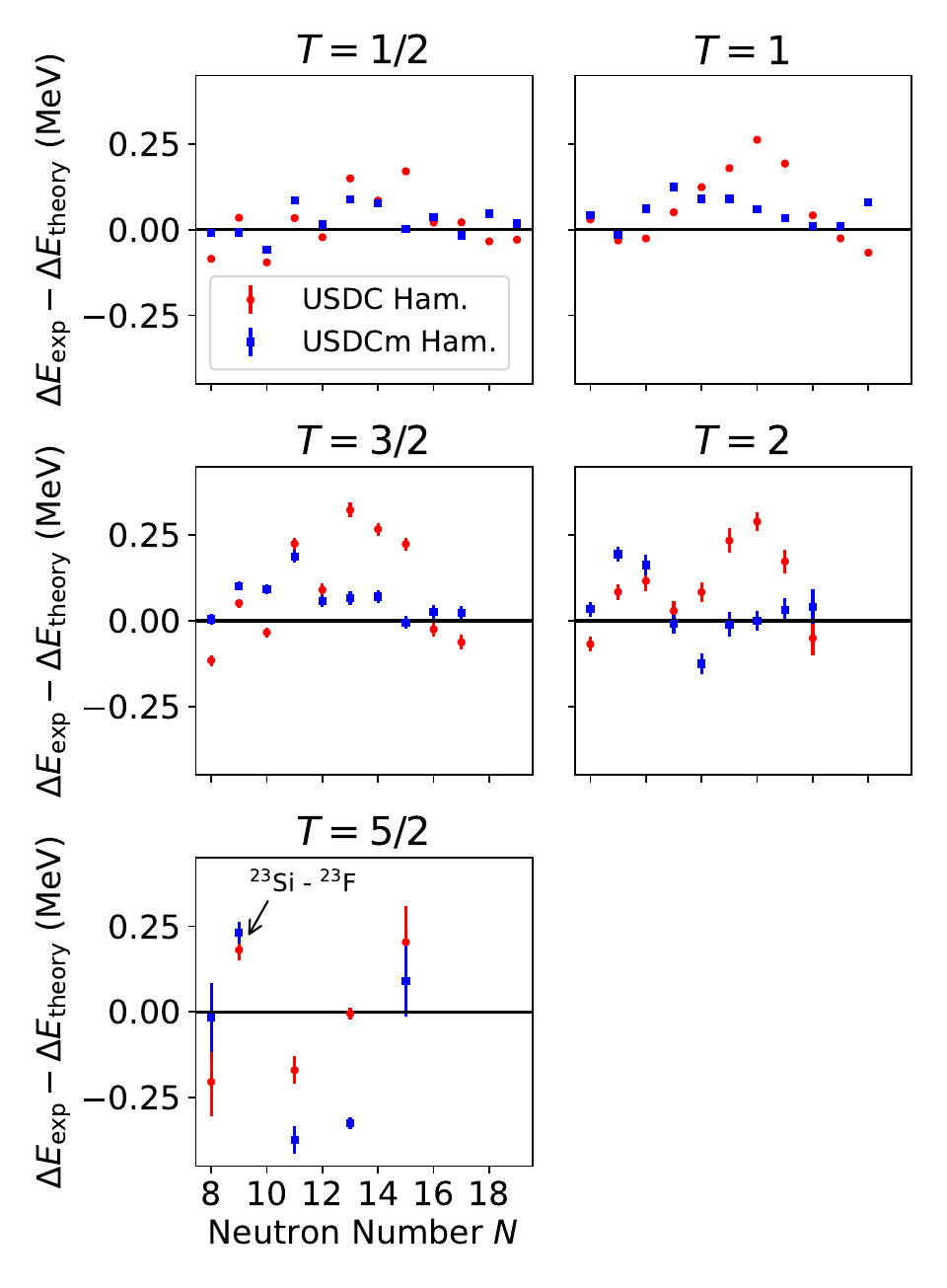}
\caption{Deviations of the binding energy differences between experiment and calculations $\Delta E_{\mathrm{exp}}-\Delta E_{\mathrm{theory}}$ as a function of the neutron number of the mirror pair nucleus with $T_z=-T$. The red points show the deviations when employing the USDC Hamiltonian and the blue squares the ones for the USDCm Hamiltonian. The error bars are only given by the respective experimental errors on the binding energies differences $\Delta E_{\mathrm{exp}}$.}
\label{fig:Deviations}
\end{figure}

When plotting the differences between the theoretical and experimental values $\Delta E_{\mathrm{exp}} - \Delta E_{\mathrm{theory}}$, see Fig.~\ref{fig:Deviations} (red dots), it can be seen that the largest deviation is typically around neutron number $N=11-15$, amounting up to 330~keV. This is associated with mirror pairs where the $1s_{1/2}$ orbital becomes occupied in the ground state. It indicates that the SPE for the $1s_{1/2}$ orbital for these cases is about 100~keV too small, which is related to the Thomas-Ehrman (TE) shift~\cite{PhysRev.81.412,PhysRev.88.1109}. The $1s_{1/2}$ SPE used for the USDC Hamiltonian is strongly influenced by the binding energy differences $\Delta E$ of excited states below $A=28$ where the $1s_{1/2}$ orbital is more loosely bound.  The valence proton in the $1s_{1/2}$ orbit of the proton-rich mirror partner is very weakly bound, leading to an increased radial extent of the wave function. Due to the reduced Coulomb repulsion, the proton-rich mirror partner is hence more bound than the neutron-rich counterpart. This results in a reduced $\Delta E$ for loosely bound states that contain some $1s_{1/2}$ proton occupation as discussed in Ref.~\cite{PhysRevC.101.064312}. 

Fig.~\ref{fig:Deviations} also shows the results for $\Delta E_{\mathrm{exp}}-\Delta E_{\mathrm{theory}}$ for another Hamiltonian called USDCm (blue squares). This USDCm Hamiltonian results by considering a fit to the $b$-coefficients and by using the single-value-decomposition method to obtain a modified set of Coulomb TBME~~\cite{PhysRevC.101.064312}. The modifications of nine linear combinations of Coulomb TBME decreased the $b$-coefficient root-mean-square deviation from 65 to 45~keV (see Fig. 6 in Ref.~\cite{PhysRevC.101.064312}). The TBME modification reduces the positive bump in the energy differences around $N=11-15$ for $T=1/2$ to $T=2$. For $T=5/2$ it however results in larger deviations for the mirror pairs $^{23}$Si - $^{23}$F,  $^{27}$S - $^{27}$Na and $^{31}$Ar - $^{31}$Al amounting up to 380~keV.

The TE shift is a nucleus and state-dependent energy shift that cannot be obtained from a fixed (universal) set of SPE and TBME. It requires one to explicitly take the nucleus and state-dependent continuum into account. The main contributing factors to the TE shift is the occupation of the $1s_{1/2}$ orbital and the average one-proton separation energy $\bar{S_{p}'}$~\cite{PhysRevC.101.064312}.
In the formalism of Ref.~\cite{PhysRevC.101.064312}, the
$  1s_{1/2}  $ orbital contribution to the TE
shift is given by
\begin{equation}
  {\rm TE}_{\mathrm{total}} = \displaystyle\sum _{E_{x}} {\rm TE}_{\mathrm{sp}}(S'_{p}) \, C^{2}S(E_{x}) , 
\label{eq: TEtotal}  
\end{equation}

where TE$_{\mathrm{sp}}$ is the single-particle TE shift given
for proton number $  Z=14  $ in Fig. 11 of Ref.~\cite{PhysRevC.101.064312} for a $^{28}$Si ($  Z=14  $) core,
and
\begin{equation}
S_{p}' = S_{p}(^{A}Z) + E_{x}[^{A-1}(Z-1)].      \label{eq:Sp'}
\end{equation}
$E_{x}[^{A-1}(Z-1)]$ is the excitation energy of the nucleus $^{A-1}(Z-1)$.
In general, there are many final states in the nucleus
$  ^{A-1}(Z-1)  $  that can be reached by the removal of a
$  1s_{1/2}  $ proton.
It is useful to consider the average excitation energy of the
final state,
\begin{equation}
\bar{E_{x}} = \frac{\displaystyle\sum _{E_{x}} E_{x}[^{A-1}(Z-1)] \,
C^{2}S(E_{x})}{\displaystyle\sum _{E_{x}} C^{2}S(E_{x})},      \label{eq:Ex}
\end{equation}
and the average proton separation energy,
\begin{equation}
\bar{S_{p}'} = S_{p}(^{A}Z) + \bar{E_{x}},     \label{eq: Sp'av}
\end{equation}
to derive an approximate expression for Eq. \ref{eq: TEtotal},
\begin{equation}
{\rm TE}^{\mathrm{ave}}_{\mathrm{total}} = {\rm TE}_{\mathrm{sp}}(\bar{S'_{p}}) \, \displaystyle\sum _{E_{x}} C^{2}S(E_{x})
= {\rm TE}_{\mathrm{sp}}(\bar{S'_{p}}) \langle 1s_{1/2}\rangle,      \label{TEtotalave}
\end{equation}
where $  \langle 1s_{1/2} \rangle  $ is the $  1s_{1/2}  $ proton occupation number.
The results for the nuclei with $  T_{z}=-5/2  $ are given in TABLE~\ref{tab: TE shift}.

\begin{table}[H] 
\centering
\caption{ Proton occupation number of the $1s_{1/2}$ orbital, theoretical separation energies and TE shifts for $T_z=-5/2$ nuclei with a neutron number $N$ between 8 and 15.}
\vspace{\baselineskip}
\begin{tabular}{l c c c c c c c }
\hline
$N$ & Nucleus & $  \langle 1s_{1/2}\rangle  $ & $  S_{p}  $ & $  \bar{S_{p}'} $  &  TE$_{\mathrm{sp}}  (\bar{S'_{p}})  $ & TE$^{\mathrm{ave}}_{\mathrm{total}}$  & {TE}$_{\mathrm{total}}$  \\ 
& & & (MeV) & (MeV) & (MeV) & (MeV) & (MeV) \\
\hline
8 & $^{21}$Al & 0.27 & -1.40  &  4.0 & -0.12 & -0.03 & -0.04\\ 
9 & $^{23}$Si & 0.52 & 2.09   &  3.8 & -0.13 & -0.07 & -0.07\\ 
10 & $^{25}$P & 1.07 & -1.57  &  1.0 & -0.32 & -0.34 & -0.88\\ 
11 & $^{27}$S & 1.52 &  0.96  &  1.4 & -0.27 & -0.41 & -0.36\\ 
12 & $^{29}$Cl& 1.75 & -2.74  &  1.0 & -0.32 & -0.56 & -0.70\\ 
13 & $^{31}$Ar& 1.78 &  0.79  &  2.2 & -0.21 & -0.37 & -0.34\\
14 & $^{33}$K & 1.95 & -2.48  &  2.7 & -0.18 & -0.35 & -0.46\\ 
15 & $^{35}$Ca& 2.00 &  1.24  &  4.4 & -0.11 & -0.22& -0.26\\
\hline
\end{tabular}
\label{tab: TE shift}
\end{table}

${\rm TE}_{\mathrm{total}}$ and ${\rm TE}^{\mathrm{ave}}_{\mathrm{total}}$ are approximately equivalent except for $^{25}$P. The reason is that the average proton separation energy amounts to $-1$~MeV, whereas the $1s_{1/2}$ removal, however, is dominated by the ground state of $^{24}$Si with a calculated ground state to ground state proton separation energy $S_p$ of -1.57 MeV and a spectroscopic factor of $C^2S=0.62$.  From Fig. 11 of Ref.~\cite{PhysRevC.101.064312}, ${\rm TE}_{\mathrm{sp}} = - 1.4$~MeV, and the contribution to the total TE shift is $-0.84$ MeV. Unfortunately, the mass of $^{25}$P is unknown up to today. Due to its short half-life of $< 30$~ns~\cite{FIRESTONE20091691} it could only be derived from decay spectroscopy. 

The masses of various other $T_z=-5/2$ nuclei were recently measured and can be compared with theory. As already discussed, Fig.~\ref{fig:Deviations} depicts the differences between the experimental and theoretical binding energy differences, $\Delta E_{\mathrm{exp}} - \Delta E_{\mathrm{theory}}$ (red points for the USDC Hamiltonian and blue squares for the USDCm Hamiltonian). 
The calculated TE shift for $^{23}$Si of $-70$~keV in Table~\ref{tab: TE shift} is small compared to the observed difference $\Delta E_{\mathrm{exp}} - \Delta E_{\mathrm{USDC}}$ of 181(31)~keV for the mirror pair $^{23}$Si - $^{23}$F . However, the deviations between experiment and theory for $T=3/2$ and $T=2$ in Fig.~\ref{fig:Deviations} are also up to 250~keV. This could be due to the use of the scaled harmonic-oscillator approximation made for the Coulomb energy calculation. Our experimental result for $^{23}$Si - $^{23}$F is consistent with the observed deviation for other nuclei. The calculated TE shift for $^{21}$Al of $-40$~keV is also small compared to the observed difference of $-204(101)$~keV for the mirror pair $^{21}$Al - $^{21}$O, but the experiment has a larger uncertainty in this case. 
For $^{27}$S and $^{31}$Ar, the $1s_{1/2}$ occupations are larger and the calculated TE shift is about -350~keV. Taking into account the theoretical uncertainty, this is consistent with the negative deviations observed for the mirror pairs $^{27}$S - $^{27}$Na and  $^{31}$Ar - $^{31}$Al in Fig.~\ref{fig:Deviations}.
The largest TE shifts are obtained for $^{25}$P and $^{29}$Cl, but the masses of these two nuclei have not yet been measured. Finally, $^{35}$Ca has a large $1s_{1/2}$ occupation, but a smaller TE shift since this state is more bound. This is consistent with the trends in Fig.~\ref{fig:Deviations}.

The generally good overall agreement between shell-model calculations~\cite{PhysRevC.101.064312} and experimental values for the binding energy differences $\Delta E$ demonstrates that the largest isospin symmetry breaking effect for isospin values up to $T=5/2$ is the Coulomb interaction.

\section{V. Conclusions}
In summary, we presented mass measurements of $^{23}$Si with a 20-fold increase in precision compared to previous measurements~\cite{PhysRevLett.133.222501}, making $^{23}$Si the nucleus with the most precisely known mass among all of the $T_z=-5/2$ nuclei.  By combining this new data with existing measurements, we investigated isospin symmetry breaking effects in $sd$-shell nuclei up to an isospin of $T=5/2$. The favorable comparison with shell-model calculations~\cite{PhysRevC.101.064312} shows that isospin symmetry breaking effects for $sd$-shell nuclei are theoretically well described, even for high isospin values up to $T=5/2$.

A day prior to the submission of this manuscript, we were made aware of an arXiv preprint~\cite{xing2025z14magicityrevealedmass} reporting another mass measurement of $^{23}$Si from CSRe. It reports a mass excess value of 23365(16)~keV, which is in excellent agreement with our value of 23362.9(5.8)~keV.  

\section{Acknowledgements}
This material is based upon work supported by the U.S. Department of Energy, Office of Science, Office of Nuclear Physics and used resources of the Facility for Rare Isotope Beams (FRIB) Operations, which is a DOE Office of Science User Facility under Award Number DE-SC0023633.
This work was conducted with the support of Michigan State University, the US National Science Foundation under contract no. PHY-2111185 and PHY-2110365, the DOE, Office of Nuclear Physics under contract no. DE-AC02-06CH11357, DE-AC02-05CH11231, DE-SC0022538, and DE-SC0022538. 
S.E.C. acknowledges support from the DOE NNSA SSGF under DE-NA0003960.

\bibliography{23Si.bib} 

\begin{thebibliography}{36}%
\makeatletter
\providecommand \@ifxundefined [1]{%
 \@ifx{#1\undefined}
}%
\providecommand \@ifnum [1]{%
 \ifnum #1\expandafter \@firstoftwo
 \else \expandafter \@secondoftwo
 \fi
}%
\providecommand \@ifx [1]{%
 \ifx #1\expandafter \@firstoftwo
 \else \expandafter \@secondoftwo
 \fi
}%
\providecommand \natexlab [1]{#1}%
\providecommand \enquote  [1]{``#1''}%
\providecommand \bibnamefont  [1]{#1}%
\providecommand \bibfnamefont [1]{#1}%
\providecommand \citenamefont [1]{#1}%
\providecommand \href@noop [0]{\@secondoftwo}%
\providecommand \href [0]{\begingroup \@sanitize@url \@href}%
\providecommand \@href[1]{\@@startlink{#1}\@@href}%
\providecommand \@@href[1]{\endgroup#1\@@endlink}%
\providecommand \@sanitize@url [0]{\catcode `\\12\catcode `\$12\catcode `\&12\catcode `\#12\catcode `\^12\catcode `\_12\catcode `\%12\relax}%
\providecommand \@@startlink[1]{}%
\providecommand \@@endlink[0]{}%
\providecommand \url  [0]{\begingroup\@sanitize@url \@url }%
\providecommand \@url [1]{\endgroup\@href {#1}{\urlprefix }}%
\providecommand \urlprefix  [0]{URL }%
\providecommand \Eprint [0]{\href }%
\providecommand \doibase [0]{https://doi.org/}%
\providecommand \selectlanguage [0]{\@gobble}%
\providecommand \bibinfo  [0]{\@secondoftwo}%
\providecommand \bibfield  [0]{\@secondoftwo}%
\providecommand \translation [1]{[#1]}%
\providecommand \BibitemOpen [0]{}%
\providecommand \bibitemStop [0]{}%
\providecommand \bibitemNoStop [0]{.\EOS\space}%
\providecommand \EOS [0]{\spacefactor3000\relax}%
\providecommand \BibitemShut  [1]{\csname bibitem#1\endcsname}%
\let\auto@bib@innerbib\@empty
\bibitem [{\citenamefont {Heisenberg}(1932)}]{Heisenberg1932}%
  \BibitemOpen
  \bibfield  {author} {\bibinfo {author} {\bibfnamefont {W.}~\bibnamefont {Heisenberg}},\ }\bibfield  {title} {\bibinfo {title} {{{\"U}ber den Bau der Atomkerne. I}},\ }\href {https://doi.org/10.1007/BF01342433} {\bibfield  {journal} {\bibinfo  {journal} {Zeitschrift f{\"u}r Physik}\ }\textbf {\bibinfo {volume} {77}},\ \bibinfo {pages} {1} (\bibinfo {year} {1932})}\BibitemShut {NoStop}%
\bibitem [{\citenamefont {Wigner}(1937)}]{PhysRev.51.106}%
  \BibitemOpen
  \bibfield  {author} {\bibinfo {author} {\bibfnamefont {E.}~\bibnamefont {Wigner}},\ }\bibfield  {title} {\bibinfo {title} {{On the Consequences of the Symmetry of the Nuclear Hamiltonian on the Spectroscopy of Nuclei}},\ }\href {https://doi.org/10.1103/PhysRev.51.106} {\bibfield  {journal} {\bibinfo  {journal} {Phys. Rev.}\ }\textbf {\bibinfo {volume} {51}},\ \bibinfo {pages} {106} (\bibinfo {year} {1937})}\BibitemShut {NoStop}%
\bibitem [{\citenamefont {Smirnova}(2023)}]{physics5020026}%
  \BibitemOpen
  \bibfield  {author} {\bibinfo {author} {\bibfnamefont {N.~A.}\ \bibnamefont {Smirnova}},\ }\bibfield  {title} {\bibinfo {title} {{Isospin-Symmetry Breaking within the Nuclear Shell Model: Present Status and Developments}},\ }\href {https://doi.org/10.3390/physics5020026} {\bibfield  {journal} {\bibinfo  {journal} {Physics}\ }\textbf {\bibinfo {volume} {5}},\ \bibinfo {pages} {352} (\bibinfo {year} {2023})}\BibitemShut {NoStop}%
\bibitem [{\citenamefont {Sheikh}\ \emph {et~al.}(2024)\citenamefont {Sheikh}, \citenamefont {Rouoof}, \citenamefont {Ali}, \citenamefont {Rather}, \citenamefont {Sarma},\ and\ \citenamefont {Srivastava}}]{sym16060745}%
  \BibitemOpen
  \bibfield  {author} {\bibinfo {author} {\bibfnamefont {J.~A.}\ \bibnamefont {Sheikh}}, \bibinfo {author} {\bibfnamefont {S.~P.}\ \bibnamefont {Rouoof}}, \bibinfo {author} {\bibfnamefont {R.~N.}\ \bibnamefont {Ali}}, \bibinfo {author} {\bibfnamefont {N.}~\bibnamefont {Rather}}, \bibinfo {author} {\bibfnamefont {C.}~\bibnamefont {Sarma}},\ and\ \bibinfo {author} {\bibfnamefont {P.~C.}\ \bibnamefont {Srivastava}},\ }\bibfield  {title} {\bibinfo {title} {{Isospin Symmetry Breaking in Atomic Nuclei}},\ }\bibfield  {journal} {\bibinfo  {journal} {Symmetry}\ }\textbf {\bibinfo {volume} {16}},\ \href {https://doi.org/10.3390/sym16060745} {10.3390/sym16060745} (\bibinfo {year} {2024})\BibitemShut {NoStop}%
\bibitem [{\citenamefont {Wrede}\ \emph {et~al.}(2009)\citenamefont {Wrede}, \citenamefont {Caggiano}, \citenamefont {Clark}, \citenamefont {Deibel}, \citenamefont {Parikh},\ and\ \citenamefont {Parker}}]{PhysRevC.79.045808}%
  \BibitemOpen
  \bibfield  {author} {\bibinfo {author} {\bibfnamefont {C.}~\bibnamefont {Wrede}}, \bibinfo {author} {\bibfnamefont {J.~A.}\ \bibnamefont {Caggiano}}, \bibinfo {author} {\bibfnamefont {J.~A.}\ \bibnamefont {Clark}}, \bibinfo {author} {\bibfnamefont {C.~M.}\ \bibnamefont {Deibel}}, \bibinfo {author} {\bibfnamefont {A.}~\bibnamefont {Parikh}},\ and\ \bibinfo {author} {\bibfnamefont {P.~D.}\ \bibnamefont {Parker}},\ }\bibfield  {title} {\bibinfo {title} {{Thermonuclear $^{30}\mathrm{S}(p,\ensuremath{\gamma})^{31}\mathrm{Cl}$ reaction in type I x-ray bursts}},\ }\href {https://doi.org/10.1103/PhysRevC.79.045808} {\bibfield  {journal} {\bibinfo  {journal} {Phys. Rev. C}\ }\textbf {\bibinfo {volume} {79}},\ \bibinfo {pages} {045808} (\bibinfo {year} {2009})}\BibitemShut {NoStop}%
\bibitem [{\citenamefont {Richter}\ and\ \citenamefont {Brown}(2013)}]{PhysRevC.87.065803}%
  \BibitemOpen
  \bibfield  {author} {\bibinfo {author} {\bibfnamefont {W.~A.}\ \bibnamefont {Richter}}\ and\ \bibinfo {author} {\bibfnamefont {B.~A.}\ \bibnamefont {Brown}},\ }\bibfield  {title} {\bibinfo {title} {{Shell-model studies of the astrophysical $\mathit{rp}$ reaction ${}^{29}$P($p,\ensuremath{\gamma}$)${}^{30}$S}},\ }\href {https://doi.org/10.1103/PhysRevC.87.065803} {\bibfield  {journal} {\bibinfo  {journal} {Phys. Rev. C}\ }\textbf {\bibinfo {volume} {87}},\ \bibinfo {pages} {065803} (\bibinfo {year} {2013})}\BibitemShut {NoStop}%
\bibitem [{\citenamefont {Ong}\ \emph {et~al.}(2017)\citenamefont {Ong}, \citenamefont {Langer}, \citenamefont {Montes}, \citenamefont {Aprahamian}, \citenamefont {Bardayan}, \citenamefont {Bazin}, \citenamefont {Brown}, \citenamefont {Browne}, \citenamefont {Crawford}, \citenamefont {Cyburt}, \citenamefont {Deleeuw}, \citenamefont {Domingo-Pardo}, \citenamefont {Gade}, \citenamefont {George}, \citenamefont {Hosmer}, \citenamefont {Keek}, \citenamefont {Kontos}, \citenamefont {Lee}, \citenamefont {Lemasson}, \citenamefont {Lunderberg}, \citenamefont {Maeda}, \citenamefont {Matos}, \citenamefont {Meisel}, \citenamefont {Noji}, \citenamefont {Nunes}, \citenamefont {Nystrom}, \citenamefont {Perdikakis}, \citenamefont {Pereira}, \citenamefont {Quinn}, \citenamefont {Recchia}, \citenamefont {Schatz}, \citenamefont {Scott}, \citenamefont {Siegl}, \citenamefont {Simon}, \citenamefont {Smith}, \citenamefont {Spyrou}, \citenamefont {Stevens}, \citenamefont {Stroberg}, \citenamefont {Weisshaar}, \citenamefont
  {Wheeler}, \citenamefont {Wimmer},\ and\ \citenamefont {Zegers}}]{PhysRevC.95.055806}%
  \BibitemOpen
  \bibfield  {author} {\bibinfo {author} {\bibfnamefont {W.-J.}\ \bibnamefont {Ong}}, \bibinfo {author} {\bibfnamefont {C.}~\bibnamefont {Langer}}, \bibinfo {author} {\bibfnamefont {F.}~\bibnamefont {Montes}}, \bibinfo {author} {\bibfnamefont {A.}~\bibnamefont {Aprahamian}}, \bibinfo {author} {\bibfnamefont {D.~W.}\ \bibnamefont {Bardayan}}, \bibinfo {author} {\bibfnamefont {D.}~\bibnamefont {Bazin}}, \bibinfo {author} {\bibfnamefont {B.~A.}\ \bibnamefont {Brown}}, \bibinfo {author} {\bibfnamefont {J.}~\bibnamefont {Browne}}, \bibinfo {author} {\bibfnamefont {H.}~\bibnamefont {Crawford}}, \bibinfo {author} {\bibfnamefont {R.}~\bibnamefont {Cyburt}}, \bibinfo {author} {\bibfnamefont {E.~B.}\ \bibnamefont {Deleeuw}}, \bibinfo {author} {\bibfnamefont {C.}~\bibnamefont {Domingo-Pardo}}, \bibinfo {author} {\bibfnamefont {A.}~\bibnamefont {Gade}}, \bibinfo {author} {\bibfnamefont {S.}~\bibnamefont {George}}, \bibinfo {author} {\bibfnamefont {P.}~\bibnamefont {Hosmer}}, \bibinfo {author} {\bibfnamefont
  {L.}~\bibnamefont {Keek}}, \bibinfo {author} {\bibfnamefont {A.}~\bibnamefont {Kontos}}, \bibinfo {author} {\bibfnamefont {I.-Y.}\ \bibnamefont {Lee}}, \bibinfo {author} {\bibfnamefont {A.}~\bibnamefont {Lemasson}}, \bibinfo {author} {\bibfnamefont {E.}~\bibnamefont {Lunderberg}}, \bibinfo {author} {\bibfnamefont {Y.}~\bibnamefont {Maeda}}, \bibinfo {author} {\bibfnamefont {M.}~\bibnamefont {Matos}}, \bibinfo {author} {\bibfnamefont {Z.}~\bibnamefont {Meisel}}, \bibinfo {author} {\bibfnamefont {S.}~\bibnamefont {Noji}}, \bibinfo {author} {\bibfnamefont {F.~M.}\ \bibnamefont {Nunes}}, \bibinfo {author} {\bibfnamefont {A.}~\bibnamefont {Nystrom}}, \bibinfo {author} {\bibfnamefont {G.}~\bibnamefont {Perdikakis}}, \bibinfo {author} {\bibfnamefont {J.}~\bibnamefont {Pereira}}, \bibinfo {author} {\bibfnamefont {S.~J.}\ \bibnamefont {Quinn}}, \bibinfo {author} {\bibfnamefont {F.}~\bibnamefont {Recchia}}, \bibinfo {author} {\bibfnamefont {H.}~\bibnamefont {Schatz}}, \bibinfo {author} {\bibfnamefont
  {M.}~\bibnamefont {Scott}}, \bibinfo {author} {\bibfnamefont {K.}~\bibnamefont {Siegl}}, \bibinfo {author} {\bibfnamefont {A.}~\bibnamefont {Simon}}, \bibinfo {author} {\bibfnamefont {M.}~\bibnamefont {Smith}}, \bibinfo {author} {\bibfnamefont {A.}~\bibnamefont {Spyrou}}, \bibinfo {author} {\bibfnamefont {J.}~\bibnamefont {Stevens}}, \bibinfo {author} {\bibfnamefont {S.~R.}\ \bibnamefont {Stroberg}}, \bibinfo {author} {\bibfnamefont {D.}~\bibnamefont {Weisshaar}}, \bibinfo {author} {\bibfnamefont {J.}~\bibnamefont {Wheeler}}, \bibinfo {author} {\bibfnamefont {K.}~\bibnamefont {Wimmer}},\ and\ \bibinfo {author} {\bibfnamefont {R.~G.~T.}\ \bibnamefont {Zegers}},\ }\bibfield  {title} {\bibinfo {title} {{Low-lying level structure of $^{56}\mathbf{Cu}$ and its implications for the $\mathit{rp}$ process}},\ }\href {https://doi.org/10.1103/PhysRevC.95.055806} {\bibfield  {journal} {\bibinfo  {journal} {Phys. Rev. C}\ }\textbf {\bibinfo {volume} {95}},\ \bibinfo {pages} {055806} (\bibinfo {year}
  {2017})}\BibitemShut {NoStop}%
\bibitem [{\citenamefont {Cirigliano}\ \emph {et~al.}(2023)\citenamefont {Cirigliano}, \citenamefont {Crivellin}, \citenamefont {Hoferichter},\ and\ \citenamefont {Moulson}}]{CIRIGLIANO2023137748}%
  \BibitemOpen
  \bibfield  {author} {\bibinfo {author} {\bibfnamefont {V.}~\bibnamefont {Cirigliano}}, \bibinfo {author} {\bibfnamefont {A.}~\bibnamefont {Crivellin}}, \bibinfo {author} {\bibfnamefont {M.}~\bibnamefont {Hoferichter}},\ and\ \bibinfo {author} {\bibfnamefont {M.}~\bibnamefont {Moulson}},\ }\bibfield  {title} {\bibinfo {title} {{Scrutinizing CKM unitarity with a new measurement of the K$\mu$3/K$\mu$2 branching fraction}},\ }\href {https://doi.org/https://doi.org/10.1016/j.physletb.2023.137748} {\bibfield  {journal} {\bibinfo  {journal} {Physics Letters B}\ }\textbf {\bibinfo {volume} {838}},\ \bibinfo {pages} {137748} (\bibinfo {year} {2023})}\BibitemShut {NoStop}%
\bibitem [{\citenamefont {Miller}\ and\ \citenamefont {Schwenk}(2008)}]{PhysRevC.78.035501}%
  \BibitemOpen
  \bibfield  {author} {\bibinfo {author} {\bibfnamefont {G.~A.}\ \bibnamefont {Miller}}\ and\ \bibinfo {author} {\bibfnamefont {A.}~\bibnamefont {Schwenk}},\ }\bibfield  {title} {\bibinfo {title} {{Isospin-symmetry-breaking corrections to superallowed Fermi $\ensuremath{\beta}$ decay: Formalism and schematic models}},\ }\href {https://doi.org/10.1103/PhysRevC.78.035501} {\bibfield  {journal} {\bibinfo  {journal} {Phys. Rev. C}\ }\textbf {\bibinfo {volume} {78}},\ \bibinfo {pages} {035501} (\bibinfo {year} {2008})}\BibitemShut {NoStop}%
\bibitem [{\citenamefont {Miller}\ and\ \citenamefont {Schwenk}(2009)}]{PhysRevC.80.064319}%
  \BibitemOpen
  \bibfield  {author} {\bibinfo {author} {\bibfnamefont {G.~A.}\ \bibnamefont {Miller}}\ and\ \bibinfo {author} {\bibfnamefont {A.}~\bibnamefont {Schwenk}},\ }\bibfield  {title} {\bibinfo {title} {{Isospin-symmetry-breaking corrections to superallowed Fermi $\ensuremath{\beta}$ decay: Radial excitations}},\ }\href {https://doi.org/10.1103/PhysRevC.80.064319} {\bibfield  {journal} {\bibinfo  {journal} {Phys. Rev. C}\ }\textbf {\bibinfo {volume} {80}},\ \bibinfo {pages} {064319} (\bibinfo {year} {2009})}\BibitemShut {NoStop}%
\bibitem [{\citenamefont {Seng}\ and\ \citenamefont {Gorchtein}(2023)}]{SENG2023137654}%
  \BibitemOpen
  \bibfield  {author} {\bibinfo {author} {\bibfnamefont {C.-Y.}\ \bibnamefont {Seng}}\ and\ \bibinfo {author} {\bibfnamefont {M.}~\bibnamefont {Gorchtein}},\ }\bibfield  {title} {\bibinfo {title} {{Electroweak nuclear radii constrain the isospin breaking correction to $V_{ud}$}},\ }\href {https://doi.org/https://doi.org/10.1016/j.physletb.2022.137654} {\bibfield  {journal} {\bibinfo  {journal} {Physics Letters B}\ }\textbf {\bibinfo {volume} {838}},\ \bibinfo {pages} {137654} (\bibinfo {year} {2023})}\BibitemShut {NoStop}%
\bibitem [{\citenamefont {Lalanne}\ \emph {et~al.}(2023)\citenamefont {Lalanne}, \citenamefont {Sorlin}, \citenamefont {Poves}, \citenamefont {Assi\'e}, \citenamefont {Hammache}, \citenamefont {Koyama}, \citenamefont {Suzuki}, \citenamefont {Flavigny}, \citenamefont {Girard-Alcindor}, \citenamefont {Lemasson}, \citenamefont {Matta}, \citenamefont {Roger}, \citenamefont {Beaumel}, \citenamefont {Blumenfeld}, \citenamefont {Brown}, \citenamefont {Santos}, \citenamefont {Delaunay}, \citenamefont {de~S\'er\'eville}, \citenamefont {Franchoo}, \citenamefont {Gibelin}, \citenamefont {Guillot}, \citenamefont {Kamalou}, \citenamefont {Kitamura}, \citenamefont {Lapoux}, \citenamefont {Mauss}, \citenamefont {Morfouace}, \citenamefont {Pancin}, \citenamefont {Saito}, \citenamefont {Stodel},\ and\ \citenamefont {Thomas}}]{PhysRevLett.131.092501}%
  \BibitemOpen
  \bibfield  {author} {\bibinfo {author} {\bibfnamefont {L.}~\bibnamefont {Lalanne}}, \bibinfo {author} {\bibfnamefont {O.}~\bibnamefont {Sorlin}}, \bibinfo {author} {\bibfnamefont {A.}~\bibnamefont {Poves}}, \bibinfo {author} {\bibfnamefont {M.}~\bibnamefont {Assi\'e}}, \bibinfo {author} {\bibfnamefont {F.}~\bibnamefont {Hammache}}, \bibinfo {author} {\bibfnamefont {S.}~\bibnamefont {Koyama}}, \bibinfo {author} {\bibfnamefont {D.}~\bibnamefont {Suzuki}}, \bibinfo {author} {\bibfnamefont {F.}~\bibnamefont {Flavigny}}, \bibinfo {author} {\bibfnamefont {V.}~\bibnamefont {Girard-Alcindor}}, \bibinfo {author} {\bibfnamefont {A.}~\bibnamefont {Lemasson}}, \bibinfo {author} {\bibfnamefont {A.}~\bibnamefont {Matta}}, \bibinfo {author} {\bibfnamefont {T.}~\bibnamefont {Roger}}, \bibinfo {author} {\bibfnamefont {D.}~\bibnamefont {Beaumel}}, \bibinfo {author} {\bibfnamefont {Y.}~\bibnamefont {Blumenfeld}}, \bibinfo {author} {\bibfnamefont {B.~A.}\ \bibnamefont {Brown}}, \bibinfo {author} {\bibfnamefont {F.~D.~O.}\
  \bibnamefont {Santos}}, \bibinfo {author} {\bibfnamefont {F.}~\bibnamefont {Delaunay}}, \bibinfo {author} {\bibfnamefont {N.}~\bibnamefont {de~S\'er\'eville}}, \bibinfo {author} {\bibfnamefont {S.}~\bibnamefont {Franchoo}}, \bibinfo {author} {\bibfnamefont {J.}~\bibnamefont {Gibelin}}, \bibinfo {author} {\bibfnamefont {J.}~\bibnamefont {Guillot}}, \bibinfo {author} {\bibfnamefont {O.}~\bibnamefont {Kamalou}}, \bibinfo {author} {\bibfnamefont {N.}~\bibnamefont {Kitamura}}, \bibinfo {author} {\bibfnamefont {V.}~\bibnamefont {Lapoux}}, \bibinfo {author} {\bibfnamefont {B.}~\bibnamefont {Mauss}}, \bibinfo {author} {\bibfnamefont {P.}~\bibnamefont {Morfouace}}, \bibinfo {author} {\bibfnamefont {J.}~\bibnamefont {Pancin}}, \bibinfo {author} {\bibfnamefont {T.~Y.}\ \bibnamefont {Saito}}, \bibinfo {author} {\bibfnamefont {C.}~\bibnamefont {Stodel}},\ and\ \bibinfo {author} {\bibfnamefont {J.-C.}\ \bibnamefont {Thomas}},\ }\bibfield  {title} {\bibinfo {title} {{$N=16$ Magicity Revealed at the Proton Drip Line
  through the Study of $^{35}\mathrm{Ca}$}},\ }\href {https://doi.org/10.1103/PhysRevLett.131.092501} {\bibfield  {journal} {\bibinfo  {journal} {Phys. Rev. Lett.}\ }\textbf {\bibinfo {volume} {131}},\ \bibinfo {pages} {092501} (\bibinfo {year} {2023})}\BibitemShut {NoStop}%
\bibitem [{\citenamefont {Kostyleva}\ \emph {et~al.}(2024)\citenamefont {Kostyleva}, \citenamefont {Xu}, \citenamefont {Mukha}, \citenamefont {Acosta}, \citenamefont {Bajzek}, \citenamefont {Casarejos}, \citenamefont {Ciemny}, \citenamefont {Cortina-Gil}, \citenamefont {Dominik}, \citenamefont {Due\~nas}, \citenamefont {Espino}, \citenamefont {Estrad\'e}, \citenamefont {Farinon}, \citenamefont {Fomichev}, \citenamefont {Geissel}, \citenamefont {Gomez-Camacho}, \citenamefont {Gorshkov}, \citenamefont {Grigorenko}, \citenamefont {Janas}, \citenamefont {Kamiski}, \citenamefont {Kiselev}, \citenamefont {Knobel}, \citenamefont {Korsheninnikov}, \citenamefont {Krupko}, \citenamefont {Kuich}, \citenamefont {Kurz}, \citenamefont {Litvinov}, \citenamefont {Marquinez-Dur\'an}, \citenamefont {Martel}, \citenamefont {Mazzocchi}, \citenamefont {Nikolskii}, \citenamefont {Nociforo}, \citenamefont {Ord\'uz}, \citenamefont {Pf\"utzner}, \citenamefont {Pietri}, \citenamefont {Pomorski}, \citenamefont {Prochazka},
  \citenamefont {Rodr\'{\i}guez-Tajes}, \citenamefont {Rymzhanova}, \citenamefont {S\'anchez-Ben\'{\i}tez}, \citenamefont {Scheidenberger}, \citenamefont {Simon}, \citenamefont {Sitar}, \citenamefont {Slepnev}, \citenamefont {Stanoiu}, \citenamefont {Strmen}, \citenamefont {S\"ummerer}, \citenamefont {Szarka}, \citenamefont {Takechi}, \citenamefont {Tanaka}, \citenamefont {Weick}, \citenamefont {Winfield}, \citenamefont {Woods},\ and\ \citenamefont {Zhukov}}]{PhysRevC.110.L031301}%
  \BibitemOpen
  \bibfield  {author} {\bibinfo {author} {\bibfnamefont {D.}~\bibnamefont {Kostyleva}}, \bibinfo {author} {\bibfnamefont {X.-D.}\ \bibnamefont {Xu}}, \bibinfo {author} {\bibfnamefont {I.}~\bibnamefont {Mukha}}, \bibinfo {author} {\bibfnamefont {L.}~\bibnamefont {Acosta}}, \bibinfo {author} {\bibfnamefont {M.}~\bibnamefont {Bajzek}}, \bibinfo {author} {\bibfnamefont {E.}~\bibnamefont {Casarejos}}, \bibinfo {author} {\bibfnamefont {A.~A.}\ \bibnamefont {Ciemny}}, \bibinfo {author} {\bibfnamefont {D.}~\bibnamefont {Cortina-Gil}}, \bibinfo {author} {\bibfnamefont {W.}~\bibnamefont {Dominik}}, \bibinfo {author} {\bibfnamefont {J.~A.}\ \bibnamefont {Due\~nas}}, \bibinfo {author} {\bibfnamefont {J.~M.}\ \bibnamefont {Espino}}, \bibinfo {author} {\bibfnamefont {A.}~\bibnamefont {Estrad\'e}}, \bibinfo {author} {\bibfnamefont {F.}~\bibnamefont {Farinon}}, \bibinfo {author} {\bibfnamefont {A.}~\bibnamefont {Fomichev}}, \bibinfo {author} {\bibfnamefont {H.}~\bibnamefont {Geissel}}, \bibinfo {author} {\bibfnamefont
  {J.}~\bibnamefont {Gomez-Camacho}}, \bibinfo {author} {\bibfnamefont {A.}~\bibnamefont {Gorshkov}}, \bibinfo {author} {\bibfnamefont {L.~V.}\ \bibnamefont {Grigorenko}}, \bibinfo {author} {\bibfnamefont {Z.}~\bibnamefont {Janas}}, \bibinfo {author} {\bibfnamefont {G.}~\bibnamefont {Kamiski}}, \bibinfo {author} {\bibfnamefont {O.}~\bibnamefont {Kiselev}}, \bibinfo {author} {\bibfnamefont {R.}~\bibnamefont {Knobel}}, \bibinfo {author} {\bibfnamefont {A.~A.}\ \bibnamefont {Korsheninnikov}}, \bibinfo {author} {\bibfnamefont {S.}~\bibnamefont {Krupko}}, \bibinfo {author} {\bibfnamefont {M.}~\bibnamefont {Kuich}}, \bibinfo {author} {\bibfnamefont {N.}~\bibnamefont {Kurz}}, \bibinfo {author} {\bibfnamefont {Y.~A.}\ \bibnamefont {Litvinov}}, \bibinfo {author} {\bibfnamefont {G.}~\bibnamefont {Marquinez-Dur\'an}}, \bibinfo {author} {\bibfnamefont {I.}~\bibnamefont {Martel}}, \bibinfo {author} {\bibfnamefont {C.}~\bibnamefont {Mazzocchi}}, \bibinfo {author} {\bibfnamefont {E.~Y.}\ \bibnamefont {Nikolskii}}, \bibinfo
  {author} {\bibfnamefont {C.}~\bibnamefont {Nociforo}}, \bibinfo {author} {\bibfnamefont {A.~K.}\ \bibnamefont {Ord\'uz}}, \bibinfo {author} {\bibfnamefont {M.}~\bibnamefont {Pf\"utzner}}, \bibinfo {author} {\bibfnamefont {S.}~\bibnamefont {Pietri}}, \bibinfo {author} {\bibfnamefont {M.}~\bibnamefont {Pomorski}}, \bibinfo {author} {\bibfnamefont {A.}~\bibnamefont {Prochazka}}, \bibinfo {author} {\bibfnamefont {C.}~\bibnamefont {Rodr\'{\i}guez-Tajes}}, \bibinfo {author} {\bibfnamefont {S.}~\bibnamefont {Rymzhanova}}, \bibinfo {author} {\bibfnamefont {A.~M.}\ \bibnamefont {S\'anchez-Ben\'{\i}tez}}, \bibinfo {author} {\bibfnamefont {C.}~\bibnamefont {Scheidenberger}}, \bibinfo {author} {\bibfnamefont {H.}~\bibnamefont {Simon}}, \bibinfo {author} {\bibfnamefont {B.}~\bibnamefont {Sitar}}, \bibinfo {author} {\bibfnamefont {R.}~\bibnamefont {Slepnev}}, \bibinfo {author} {\bibfnamefont {M.}~\bibnamefont {Stanoiu}}, \bibinfo {author} {\bibfnamefont {P.}~\bibnamefont {Strmen}}, \bibinfo {author} {\bibfnamefont
  {K.}~\bibnamefont {S\"ummerer}}, \bibinfo {author} {\bibfnamefont {I.}~\bibnamefont {Szarka}}, \bibinfo {author} {\bibfnamefont {M.}~\bibnamefont {Takechi}}, \bibinfo {author} {\bibfnamefont {Y.~K.}\ \bibnamefont {Tanaka}}, \bibinfo {author} {\bibfnamefont {H.}~\bibnamefont {Weick}}, \bibinfo {author} {\bibfnamefont {J.~S.}\ \bibnamefont {Winfield}}, \bibinfo {author} {\bibfnamefont {P.~J.}\ \bibnamefont {Woods}},\ and\ \bibinfo {author} {\bibfnamefont {M.~V.}\ \bibnamefont {Zhukov}},\ }\bibfield  {title} {\bibinfo {title} {{Observation and spectroscopy of the proton-unbound nucleus $^{21}\mathrm{Al}$}},\ }\href {https://doi.org/10.1103/PhysRevC.110.L031301} {\bibfield  {journal} {\bibinfo  {journal} {Phys. Rev. C}\ }\textbf {\bibinfo {volume} {110}},\ \bibinfo {pages} {L031301} (\bibinfo {year} {2024})}\BibitemShut {NoStop}%
\bibitem [{\citenamefont {Yu}\ \emph {et~al.}(2024)\citenamefont {Yu}, \citenamefont {Xing}, \citenamefont {Zhang}, \citenamefont {Wang}, \citenamefont {Zhou}, \citenamefont {Li}, \citenamefont {Li}, \citenamefont {Yuan}, \citenamefont {Niu}, \citenamefont {Huang}, \citenamefont {Geng}, \citenamefont {Guo}, \citenamefont {Chen}, \citenamefont {Pei}, \citenamefont {Xu}, \citenamefont {Litvinov}, \citenamefont {Blaum}, \citenamefont {de~Angelis}, \citenamefont {Tanihata}, \citenamefont {Yamaguchi}, \citenamefont {Zhou}, \citenamefont {Xu}, \citenamefont {Chen}, \citenamefont {Chen}, \citenamefont {Deng}, \citenamefont {Fu}, \citenamefont {Ge}, \citenamefont {Huang}, \citenamefont {Jiao}, \citenamefont {Luo}, \citenamefont {Li}, \citenamefont {Liao}, \citenamefont {Shi}, \citenamefont {Si}, \citenamefont {Sun}, \citenamefont {Shuai}, \citenamefont {Tu}, \citenamefont {Wang}, \citenamefont {Xu}, \citenamefont {Yan}, \citenamefont {Yuan},\ and\ \citenamefont {Zhang}}]{PhysRevLett.133.222501}%
  \BibitemOpen
  \bibfield  {author} {\bibinfo {author} {\bibfnamefont {Y.}~\bibnamefont {Yu}}, \bibinfo {author} {\bibfnamefont {Y.~M.}\ \bibnamefont {Xing}}, \bibinfo {author} {\bibfnamefont {Y.~H.}\ \bibnamefont {Zhang}}, \bibinfo {author} {\bibfnamefont {M.}~\bibnamefont {Wang}}, \bibinfo {author} {\bibfnamefont {X.~H.}\ \bibnamefont {Zhou}}, \bibinfo {author} {\bibfnamefont {J.~G.}\ \bibnamefont {Li}}, \bibinfo {author} {\bibfnamefont {H.~H.}\ \bibnamefont {Li}}, \bibinfo {author} {\bibfnamefont {Q.}~\bibnamefont {Yuan}}, \bibinfo {author} {\bibfnamefont {Y.~F.}\ \bibnamefont {Niu}}, \bibinfo {author} {\bibfnamefont {Y.~N.}\ \bibnamefont {Huang}}, \bibinfo {author} {\bibfnamefont {J.}~\bibnamefont {Geng}}, \bibinfo {author} {\bibfnamefont {J.~Y.}\ \bibnamefont {Guo}}, \bibinfo {author} {\bibfnamefont {J.~W.}\ \bibnamefont {Chen}}, \bibinfo {author} {\bibfnamefont {J.~C.}\ \bibnamefont {Pei}}, \bibinfo {author} {\bibfnamefont {F.~R.}\ \bibnamefont {Xu}}, \bibinfo {author} {\bibfnamefont {Y.~A.}\ \bibnamefont
  {Litvinov}}, \bibinfo {author} {\bibfnamefont {K.}~\bibnamefont {Blaum}}, \bibinfo {author} {\bibfnamefont {G.}~\bibnamefont {de~Angelis}}, \bibinfo {author} {\bibfnamefont {I.}~\bibnamefont {Tanihata}}, \bibinfo {author} {\bibfnamefont {T.}~\bibnamefont {Yamaguchi}}, \bibinfo {author} {\bibfnamefont {X.}~\bibnamefont {Zhou}}, \bibinfo {author} {\bibfnamefont {H.~S.}\ \bibnamefont {Xu}}, \bibinfo {author} {\bibfnamefont {Z.~Y.}\ \bibnamefont {Chen}}, \bibinfo {author} {\bibfnamefont {R.~J.}\ \bibnamefont {Chen}}, \bibinfo {author} {\bibfnamefont {H.~Y.}\ \bibnamefont {Deng}}, \bibinfo {author} {\bibfnamefont {C.~Y.}\ \bibnamefont {Fu}}, \bibinfo {author} {\bibfnamefont {W.~W.}\ \bibnamefont {Ge}}, \bibinfo {author} {\bibfnamefont {W.~J.}\ \bibnamefont {Huang}}, \bibinfo {author} {\bibfnamefont {H.~Y.}\ \bibnamefont {Jiao}}, \bibinfo {author} {\bibfnamefont {Y.~F.}\ \bibnamefont {Luo}}, \bibinfo {author} {\bibfnamefont {H.~F.}\ \bibnamefont {Li}}, \bibinfo {author} {\bibfnamefont {T.}~\bibnamefont {Liao}},
  \bibinfo {author} {\bibfnamefont {J.~Y.}\ \bibnamefont {Shi}}, \bibinfo {author} {\bibfnamefont {M.}~\bibnamefont {Si}}, \bibinfo {author} {\bibfnamefont {M.~Z.}\ \bibnamefont {Sun}}, \bibinfo {author} {\bibfnamefont {P.}~\bibnamefont {Shuai}}, \bibinfo {author} {\bibfnamefont {X.~L.}\ \bibnamefont {Tu}}, \bibinfo {author} {\bibfnamefont {Q.}~\bibnamefont {Wang}}, \bibinfo {author} {\bibfnamefont {X.}~\bibnamefont {Xu}}, \bibinfo {author} {\bibfnamefont {X.~L.}\ \bibnamefont {Yan}}, \bibinfo {author} {\bibfnamefont {Y.~J.}\ \bibnamefont {Yuan}},\ and\ \bibinfo {author} {\bibfnamefont {M.}~\bibnamefont {Zhang}},\ }\bibfield  {title} {\bibinfo {title} {{Nuclear Structure of Dripline Nuclei Elucidated through Precision Mass Measurements of $^{23}\mathrm{Si}$, $^{26}\mathrm{P}$, $^{27,28}\mathrm{S}$, and $^{31}\mathrm{Ar}$}},\ }\href {https://doi.org/10.1103/PhysRevLett.133.222501} {\bibfield  {journal} {\bibinfo  {journal} {Phys. Rev. Lett.}\ }\textbf {\bibinfo {volume} {133}},\ \bibinfo {pages} {222501}
  (\bibinfo {year} {2024})}\BibitemShut {NoStop}%
\bibitem [{\citenamefont {Magilligan}\ and\ \citenamefont {Brown}(2020)}]{PhysRevC.101.064312}%
  \BibitemOpen
  \bibfield  {author} {\bibinfo {author} {\bibfnamefont {A.}~\bibnamefont {Magilligan}}\ and\ \bibinfo {author} {\bibfnamefont {B.~A.}\ \bibnamefont {Brown}},\ }\bibfield  {title} {\bibinfo {title} {{New isospin-breaking ``USD'' Hamiltonians for the sd shell}},\ }\href {https://doi.org/10.1103/PhysRevC.101.064312} {\bibfield  {journal} {\bibinfo  {journal} {Phys. Rev. C}\ }\textbf {\bibinfo {volume} {101}},\ \bibinfo {pages} {064312} (\bibinfo {year} {2020})}\BibitemShut {NoStop}%
\bibitem [{\citenamefont {Wang}\ \emph {et~al.}(2021)\citenamefont {Wang}, \citenamefont {Huang}, \citenamefont {Kondev}, \citenamefont {Audi},\ and\ \citenamefont {Naimi}}]{AME2020}%
  \BibitemOpen
  \bibfield  {author} {\bibinfo {author} {\bibfnamefont {M.}~\bibnamefont {Wang}}, \bibinfo {author} {\bibfnamefont {W.}~\bibnamefont {Huang}}, \bibinfo {author} {\bibfnamefont {F.}~\bibnamefont {Kondev}}, \bibinfo {author} {\bibfnamefont {G.}~\bibnamefont {Audi}},\ and\ \bibinfo {author} {\bibfnamefont {S.}~\bibnamefont {Naimi}},\ }\bibfield  {title} {\bibinfo {title} {{The AME 2020 atomic mass evaluation (II). Tables, graphs and references*}},\ }\href {https://doi.org/10.1088/1674-1137/abddaf} {\bibfield  {journal} {\bibinfo  {journal} {Chinese Physics C}\ }\textbf {\bibinfo {volume} {45}},\ \bibinfo {pages} {030003} (\bibinfo {year} {2021})}\BibitemShut {NoStop}%
\bibitem [{\citenamefont {Campbell}\ \emph {et~al.}(2024)\citenamefont {Campbell}, \citenamefont {Bollen}, \citenamefont {Brown}, \citenamefont {Dockery}, \citenamefont {Ireland}, \citenamefont {Minamisono}, \citenamefont {Puentes}, \citenamefont {Rickey}, \citenamefont {Ringle}, \citenamefont {Yandow}, \citenamefont {Fossez}, \citenamefont {Ortiz-Cortes}, \citenamefont {Schwarz}, \citenamefont {Sumithrarachchi},\ and\ \citenamefont {Villari}}]{PhysRevLett.132.152501}%
  \BibitemOpen
  \bibfield  {author} {\bibinfo {author} {\bibfnamefont {S.~E.}\ \bibnamefont {Campbell}}, \bibinfo {author} {\bibfnamefont {G.}~\bibnamefont {Bollen}}, \bibinfo {author} {\bibfnamefont {B.~A.}\ \bibnamefont {Brown}}, \bibinfo {author} {\bibfnamefont {A.}~\bibnamefont {Dockery}}, \bibinfo {author} {\bibfnamefont {C.~M.}\ \bibnamefont {Ireland}}, \bibinfo {author} {\bibfnamefont {K.}~\bibnamefont {Minamisono}}, \bibinfo {author} {\bibfnamefont {D.}~\bibnamefont {Puentes}}, \bibinfo {author} {\bibfnamefont {B.~J.}\ \bibnamefont {Rickey}}, \bibinfo {author} {\bibfnamefont {R.}~\bibnamefont {Ringle}}, \bibinfo {author} {\bibfnamefont {I.~T.}\ \bibnamefont {Yandow}}, \bibinfo {author} {\bibfnamefont {K.}~\bibnamefont {Fossez}}, \bibinfo {author} {\bibfnamefont {A.}~\bibnamefont {Ortiz-Cortes}}, \bibinfo {author} {\bibfnamefont {S.}~\bibnamefont {Schwarz}}, \bibinfo {author} {\bibfnamefont {C.~S.}\ \bibnamefont {Sumithrarachchi}},\ and\ \bibinfo {author} {\bibfnamefont {A.~C.~C.}\ \bibnamefont {Villari}},\
  }\bibfield  {title} {\bibinfo {title} {{Precision Mass Measurement of the Proton Dripline Halo Candidate $^{22}\mathrm{Al}$}},\ }\href {https://doi.org/10.1103/PhysRevLett.132.152501} {\bibfield  {journal} {\bibinfo  {journal} {Phys. Rev. Lett.}\ }\textbf {\bibinfo {volume} {132}},\ \bibinfo {pages} {152501} (\bibinfo {year} {2024})}\BibitemShut {NoStop}%
\bibitem [{\citenamefont {Dronchi}\ \emph {et~al.}(2024)\citenamefont {Dronchi}, \citenamefont {Charity}, \citenamefont {Sobotka}, \citenamefont {Brown}, \citenamefont {Weisshaar}, \citenamefont {Gade}, \citenamefont {Brown}, \citenamefont {Reviol}, \citenamefont {Bazin}, \citenamefont {Farris}, \citenamefont {Hill}, \citenamefont {Li}, \citenamefont {Longfellow}, \citenamefont {Rhodes}, \citenamefont {Paneru}, \citenamefont {Gillespie}, \citenamefont {Anthony}, \citenamefont {Rubino},\ and\ \citenamefont {Biswas}}]{PhysRevC.110.L031302}%
  \BibitemOpen
  \bibfield  {author} {\bibinfo {author} {\bibfnamefont {N.}~\bibnamefont {Dronchi}}, \bibinfo {author} {\bibfnamefont {R.~J.}\ \bibnamefont {Charity}}, \bibinfo {author} {\bibfnamefont {L.~G.}\ \bibnamefont {Sobotka}}, \bibinfo {author} {\bibfnamefont {B.~A.}\ \bibnamefont {Brown}}, \bibinfo {author} {\bibfnamefont {D.}~\bibnamefont {Weisshaar}}, \bibinfo {author} {\bibfnamefont {A.}~\bibnamefont {Gade}}, \bibinfo {author} {\bibfnamefont {K.~W.}\ \bibnamefont {Brown}}, \bibinfo {author} {\bibfnamefont {W.}~\bibnamefont {Reviol}}, \bibinfo {author} {\bibfnamefont {D.}~\bibnamefont {Bazin}}, \bibinfo {author} {\bibfnamefont {P.~J.}\ \bibnamefont {Farris}}, \bibinfo {author} {\bibfnamefont {A.~M.}\ \bibnamefont {Hill}}, \bibinfo {author} {\bibfnamefont {J.}~\bibnamefont {Li}}, \bibinfo {author} {\bibfnamefont {B.}~\bibnamefont {Longfellow}}, \bibinfo {author} {\bibfnamefont {D.}~\bibnamefont {Rhodes}}, \bibinfo {author} {\bibfnamefont {S.~N.}\ \bibnamefont {Paneru}}, \bibinfo {author} {\bibfnamefont {S.~A.}\
  \bibnamefont {Gillespie}}, \bibinfo {author} {\bibfnamefont {A.~K.}\ \bibnamefont {Anthony}}, \bibinfo {author} {\bibfnamefont {E.}~\bibnamefont {Rubino}},\ and\ \bibinfo {author} {\bibfnamefont {S.}~\bibnamefont {Biswas}},\ }\bibfield  {title} {\bibinfo {title} {{Evolution of shell gaps in the neutron-poor calcium region from invariant-mass spectroscopy of $^{37,38}\mathrm{Sc}, ^{35}\mathrm{Ca}$, and $^{34}\mathrm{K}$}},\ }\href {https://doi.org/10.1103/PhysRevC.110.L031302} {\bibfield  {journal} {\bibinfo  {journal} {Phys. Rev. C}\ }\textbf {\bibinfo {volume} {110}},\ \bibinfo {pages} {L031302} (\bibinfo {year} {2024})}\BibitemShut {NoStop}%
\bibitem [{\citenamefont {York}\ \emph {et~al.}(2009)\citenamefont {York}, \citenamefont {Bollen}, \citenamefont {Compton}, \citenamefont {Crawford}, \citenamefont {Doleans}, \citenamefont {Glasmacher}, \citenamefont {Hartung}, \citenamefont {Marti}, \citenamefont {Popielarski}, \citenamefont {Vincent} \emph {et~al.}}]{York-LINAC}%
  \BibitemOpen
  \bibfield  {author} {\bibinfo {author} {\bibfnamefont {R.}~\bibnamefont {York}}, \bibinfo {author} {\bibfnamefont {G.}~\bibnamefont {Bollen}}, \bibinfo {author} {\bibfnamefont {C.}~\bibnamefont {Compton}}, \bibinfo {author} {\bibfnamefont {A.}~\bibnamefont {Crawford}}, \bibinfo {author} {\bibfnamefont {M.}~\bibnamefont {Doleans}}, \bibinfo {author} {\bibfnamefont {T.}~\bibnamefont {Glasmacher}}, \bibinfo {author} {\bibfnamefont {W.}~\bibnamefont {Hartung}}, \bibinfo {author} {\bibfnamefont {F.}~\bibnamefont {Marti}}, \bibinfo {author} {\bibfnamefont {J.}~\bibnamefont {Popielarski}}, \bibinfo {author} {\bibfnamefont {J.}~\bibnamefont {Vincent}}, \emph {et~al.},\ }\bibfield  {title} {\bibinfo {title} {{FRIB: A new accelerator facility for the production of rare isotope beams}},\ }\href@noop {} {\bibfield  {journal} {\bibinfo  {journal} {SRF2009, Berlin, September}\ } (\bibinfo {year} {2009})}\BibitemShut {NoStop}%
\bibitem [{\citenamefont {Hausmann}\ \emph {et~al.}(2013)\citenamefont {Hausmann}, \citenamefont {Aaron}, \citenamefont {Amthor}, \citenamefont {Avilov}, \citenamefont {Bandura}, \citenamefont {Bennett}, \citenamefont {Bollen}, \citenamefont {Borden}, \citenamefont {Burgess}, \citenamefont {Chouhan} \emph {et~al.}}]{Hausmann-ARIS}%
  \BibitemOpen
  \bibfield  {author} {\bibinfo {author} {\bibfnamefont {M.}~\bibnamefont {Hausmann}}, \bibinfo {author} {\bibfnamefont {A.}~\bibnamefont {Aaron}}, \bibinfo {author} {\bibfnamefont {A.}~\bibnamefont {Amthor}}, \bibinfo {author} {\bibfnamefont {M.}~\bibnamefont {Avilov}}, \bibinfo {author} {\bibfnamefont {L.}~\bibnamefont {Bandura}}, \bibinfo {author} {\bibfnamefont {R.}~\bibnamefont {Bennett}}, \bibinfo {author} {\bibfnamefont {G.}~\bibnamefont {Bollen}}, \bibinfo {author} {\bibfnamefont {T.}~\bibnamefont {Borden}}, \bibinfo {author} {\bibfnamefont {T.}~\bibnamefont {Burgess}}, \bibinfo {author} {\bibfnamefont {S.}~\bibnamefont {Chouhan}}, \emph {et~al.},\ }\bibfield  {title} {\bibinfo {title} {{Design of the advanced rare isotope separator ARIS at FRIB}},\ }\href@noop {} {\bibfield  {journal} {\bibinfo  {journal} {Nuclear Instruments and Methods in Physics Research Section B: Beam Interactions with Materials and Atoms}\ }\textbf {\bibinfo {volume} {317}},\ \bibinfo {pages} {349} (\bibinfo {year}
  {2013})}\BibitemShut {NoStop}%
\bibitem [{\citenamefont {Lund}\ \emph {et~al.}(2020)\citenamefont {Lund}, \citenamefont {Bollen}, \citenamefont {Lawton}, \citenamefont {Morrissey}, \citenamefont {Ottarson}, \citenamefont {Ringle}, \citenamefont {Schwarz}, \citenamefont {Sumithrarachchi}, \citenamefont {Villari},\ and\ \citenamefont {Yurkon}}]{Lund-ACGS}%
  \BibitemOpen
  \bibfield  {author} {\bibinfo {author} {\bibfnamefont {K.}~\bibnamefont {Lund}}, \bibinfo {author} {\bibfnamefont {G.}~\bibnamefont {Bollen}}, \bibinfo {author} {\bibfnamefont {D.}~\bibnamefont {Lawton}}, \bibinfo {author} {\bibfnamefont {D.}~\bibnamefont {Morrissey}}, \bibinfo {author} {\bibfnamefont {J.}~\bibnamefont {Ottarson}}, \bibinfo {author} {\bibfnamefont {R.}~\bibnamefont {Ringle}}, \bibinfo {author} {\bibfnamefont {S.}~\bibnamefont {Schwarz}}, \bibinfo {author} {\bibfnamefont {C.}~\bibnamefont {Sumithrarachchi}}, \bibinfo {author} {\bibfnamefont {A.}~\bibnamefont {Villari}},\ and\ \bibinfo {author} {\bibfnamefont {J.}~\bibnamefont {Yurkon}},\ }\bibfield  {title} {\bibinfo {title} {{Online tests of the advanced cryogenic gas stopper at NSCL}},\ }\href@noop {} {\bibfield  {journal} {\bibinfo  {journal} {Nuclear Instruments and Methods in Physics Research Section B: Beam Interactions with Materials and Atoms}\ }\textbf {\bibinfo {volume} {463}},\ \bibinfo {pages} {378} (\bibinfo {year}
  {2020})}\BibitemShut {NoStop}%
\bibitem [{\citenamefont {Bollen}(2011)}]{BOLLEN-IonSurfing}%
  \BibitemOpen
  \bibfield  {author} {\bibinfo {author} {\bibfnamefont {G.}~\bibnamefont {Bollen}},\ }\bibfield  {title} {\bibinfo {title} {{“Ion surfing” with radiofrequency carpets}},\ }\href {https://doi.org/https://doi.org/10.1016/j.ijms.2010.09.032} {\bibfield  {journal} {\bibinfo  {journal} {International Journal of Mass Spectrometry}\ }\textbf {\bibinfo {volume} {299}},\ \bibinfo {pages} {131} (\bibinfo {year} {2011})}\BibitemShut {NoStop}%
\bibitem [{\citenamefont {Schwarz}\ \emph {et~al.}(2016)\citenamefont {Schwarz}, \citenamefont {Bollen}, \citenamefont {Ringle}, \citenamefont {Savory},\ and\ \citenamefont {Schury}}]{CoolerBuncher}%
  \BibitemOpen
  \bibfield  {author} {\bibinfo {author} {\bibfnamefont {S.}~\bibnamefont {Schwarz}}, \bibinfo {author} {\bibfnamefont {G.}~\bibnamefont {Bollen}}, \bibinfo {author} {\bibfnamefont {R.}~\bibnamefont {Ringle}}, \bibinfo {author} {\bibfnamefont {J.}~\bibnamefont {Savory}},\ and\ \bibinfo {author} {\bibfnamefont {P.}~\bibnamefont {Schury}},\ }\bibfield  {title} {\bibinfo {title} {{The LEBIT ion cooler and buncher}},\ }\href {https://doi.org/https://doi.org/10.1016/j.nima.2016.01.078} {\bibfield  {journal} {\bibinfo  {journal} {Nuclear Instruments and Methods in Physics Research Section A: Accelerators, Spectrometers, Detectors and Associated Equipment}\ }\textbf {\bibinfo {volume} {816}},\ \bibinfo {pages} {131} (\bibinfo {year} {2016})}\BibitemShut {NoStop}%
\bibitem [{\citenamefont {Ringle}\ \emph {et~al.}(2009)\citenamefont {Ringle}, \citenamefont {Bollen}, \citenamefont {Prinke}, \citenamefont {Savory}, \citenamefont {Schury}, \citenamefont {Schwarz},\ and\ \citenamefont {Sun}}]{PenningTrap}%
  \BibitemOpen
  \bibfield  {author} {\bibinfo {author} {\bibfnamefont {R.}~\bibnamefont {Ringle}}, \bibinfo {author} {\bibfnamefont {G.}~\bibnamefont {Bollen}}, \bibinfo {author} {\bibfnamefont {A.}~\bibnamefont {Prinke}}, \bibinfo {author} {\bibfnamefont {J.}~\bibnamefont {Savory}}, \bibinfo {author} {\bibfnamefont {P.}~\bibnamefont {Schury}}, \bibinfo {author} {\bibfnamefont {S.}~\bibnamefont {Schwarz}},\ and\ \bibinfo {author} {\bibfnamefont {T.}~\bibnamefont {Sun}},\ }\bibfield  {title} {\bibinfo {title} {{The LEBIT 9.4T Penning trap mass spectrometer}},\ }\href {https://doi.org/https://doi.org/10.1016/j.nima.2009.03.207} {\bibfield  {journal} {\bibinfo  {journal} {Nuclear Instruments and Methods in Physics Research Section A: Accelerators, Spectrometers, Detectors and Associated Equipment}\ }\textbf {\bibinfo {volume} {604}},\ \bibinfo {pages} {536} (\bibinfo {year} {2009})}\BibitemShut {NoStop}%
\bibitem [{\citenamefont {Gabrielse}(2009)}]{Gabrielse-Sideband}%
  \BibitemOpen
  \bibfield  {author} {\bibinfo {author} {\bibfnamefont {G.}~\bibnamefont {Gabrielse}},\ }\bibfield  {title} {\bibinfo {title} {{Why is sideband mass spectrometry possible with ions in a Penning trap?}},\ }\href@noop {} {\bibfield  {journal} {\bibinfo  {journal} {Phys. Rev. Lett.}\ }\textbf {\bibinfo {volume} {102}},\ \bibinfo {pages} {172501} (\bibinfo {year} {2009})}\BibitemShut {NoStop}%
\bibitem [{\citenamefont {Bollen}\ \emph {et~al.}(1990)\citenamefont {Bollen}, \citenamefont {Moore}, \citenamefont {Savard},\ and\ \citenamefont {Stolzenberg}}]{ToF1}%
  \BibitemOpen
  \bibfield  {author} {\bibinfo {author} {\bibfnamefont {G.}~\bibnamefont {Bollen}}, \bibinfo {author} {\bibfnamefont {R.~B.}\ \bibnamefont {Moore}}, \bibinfo {author} {\bibfnamefont {G.}~\bibnamefont {Savard}},\ and\ \bibinfo {author} {\bibfnamefont {H.}~\bibnamefont {Stolzenberg}},\ }\bibfield  {title} {\bibinfo {title} {{The accuracy of heavy‐ion mass measurements using time of flight‐ion cyclotron resonance in a Penning trap}},\ }\href {https://doi.org/10.1063/1.346185} {\bibfield  {journal} {\bibinfo  {journal} {Journal of Applied Physics}\ }\textbf {\bibinfo {volume} {68}},\ \bibinfo {pages} {4355} (\bibinfo {year} {1990})},\ \Eprint {https://arxiv.org/abs/https://doi.org/10.1063/1.346185} {https://doi.org/10.1063/1.346185} \BibitemShut {NoStop}%
\bibitem [{\citenamefont {Becker}\ \emph {et~al.}(1990)\citenamefont {Becker}, \citenamefont {Bollen}, \citenamefont {Kern}, \citenamefont {Kluge}, \citenamefont {Moore}, \citenamefont {Savard}, \citenamefont {Schweikhard},\ and\ \citenamefont {Stolzenberg}}]{BECKER199053}%
  \BibitemOpen
  \bibfield  {author} {\bibinfo {author} {\bibfnamefont {S.}~\bibnamefont {Becker}}, \bibinfo {author} {\bibfnamefont {G.}~\bibnamefont {Bollen}}, \bibinfo {author} {\bibfnamefont {F.}~\bibnamefont {Kern}}, \bibinfo {author} {\bibfnamefont {H.-J.}\ \bibnamefont {Kluge}}, \bibinfo {author} {\bibfnamefont {R.}~\bibnamefont {Moore}}, \bibinfo {author} {\bibfnamefont {G.}~\bibnamefont {Savard}}, \bibinfo {author} {\bibfnamefont {L.}~\bibnamefont {Schweikhard}},\ and\ \bibinfo {author} {\bibfnamefont {H.}~\bibnamefont {Stolzenberg}},\ }\bibfield  {title} {\bibinfo {title} {{Mass measurements of very high accuracy by time-of-flight ion cyclotron resonance of ions injected into a Penning trap}},\ }\href {https://doi.org/https://doi.org/10.1016/0168-1176(90)85021-S} {\bibfield  {journal} {\bibinfo  {journal} {International Journal of Mass Spectrometry and Ion Processes}\ }\textbf {\bibinfo {volume} {99}},\ \bibinfo {pages} {53} (\bibinfo {year} {1990})}\BibitemShut {NoStop}%
\bibitem [{\citenamefont {König}\ \emph {et~al.}(1995)\citenamefont {König}, \citenamefont {Bollen}, \citenamefont {Kluge}, \citenamefont {Otto},\ and\ \citenamefont {Szerypo}}]{KONIG199595}%
  \BibitemOpen
  \bibfield  {author} {\bibinfo {author} {\bibfnamefont {M.}~\bibnamefont {König}}, \bibinfo {author} {\bibfnamefont {G.}~\bibnamefont {Bollen}}, \bibinfo {author} {\bibfnamefont {H.-J.}\ \bibnamefont {Kluge}}, \bibinfo {author} {\bibfnamefont {T.}~\bibnamefont {Otto}},\ and\ \bibinfo {author} {\bibfnamefont {J.}~\bibnamefont {Szerypo}},\ }\bibfield  {title} {\bibinfo {title} {{Quadrupole excitation of stored ion motion at the true cyclotron frequency}},\ }\href {https://doi.org/https://doi.org/10.1016/0168-1176(95)04146-C} {\bibfield  {journal} {\bibinfo  {journal} {International Journal of Mass Spectrometry and Ion Processes}\ }\textbf {\bibinfo {volume} {142}},\ \bibinfo {pages} {95} (\bibinfo {year} {1995})}\BibitemShut {NoStop}%
\bibitem [{\citenamefont {Ringle}\ \emph {et~al.}(2007{\natexlab{a}})\citenamefont {Ringle}, \citenamefont {Bollen}, \citenamefont {Prinke}, \citenamefont {Savory}, \citenamefont {Schury}, \citenamefont {Schwarz},\ and\ \citenamefont {Sun}}]{LorentzSteerer}%
  \BibitemOpen
  \bibfield  {author} {\bibinfo {author} {\bibfnamefont {R.}~\bibnamefont {Ringle}}, \bibinfo {author} {\bibfnamefont {G.}~\bibnamefont {Bollen}}, \bibinfo {author} {\bibfnamefont {A.}~\bibnamefont {Prinke}}, \bibinfo {author} {\bibfnamefont {J.}~\bibnamefont {Savory}}, \bibinfo {author} {\bibfnamefont {P.}~\bibnamefont {Schury}}, \bibinfo {author} {\bibfnamefont {S.}~\bibnamefont {Schwarz}},\ and\ \bibinfo {author} {\bibfnamefont {T.}~\bibnamefont {Sun}},\ }\bibfield  {title} {\bibinfo {title} {{A “Lorentz” steerer for ion injection into a Penning trap}},\ }\href {https://doi.org/https://doi.org/10.1016/j.ijms.2006.12.008} {\bibfield  {journal} {\bibinfo  {journal} {International Journal of Mass Spectrometry}\ }\textbf {\bibinfo {volume} {263}},\ \bibinfo {pages} {38} (\bibinfo {year} {2007}{\natexlab{a}})}\BibitemShut {NoStop}%
\bibitem [{\citenamefont {{Shamsuzzoha Basunia}}\ and\ \citenamefont {Chakraborty}(2021)}]{SHAMSUZZOHABASUNIA20211}%
  \BibitemOpen
  \bibfield  {author} {\bibinfo {author} {\bibfnamefont {M.}~\bibnamefont {{Shamsuzzoha Basunia}}}\ and\ \bibinfo {author} {\bibfnamefont {A.}~\bibnamefont {Chakraborty}},\ }\bibfield  {title} {\bibinfo {title} {{Nuclear Data Sheets for A=23}},\ }\href {https://doi.org/https://doi.org/10.1016/j.nds.2020.12.001} {\bibfield  {journal} {\bibinfo  {journal} {Nuclear Data Sheets}\ }\textbf {\bibinfo {volume} {171}},\ \bibinfo {pages} {1} (\bibinfo {year} {2021})}\BibitemShut {NoStop}%
\bibitem [{\citenamefont {Gulyuz}\ \emph {et~al.}(2015)\citenamefont {Gulyuz}, \citenamefont {Ariche}, \citenamefont {Bollen}, \citenamefont {Bustabad}, \citenamefont {Eibach}, \citenamefont {Izzo}, \citenamefont {Novario}, \citenamefont {Redshaw}, \citenamefont {Ringle}, \citenamefont {Sandler}, \citenamefont {Schwarz},\ and\ \citenamefont {Valverde}}]{MassOffsetError}%
  \BibitemOpen
  \bibfield  {author} {\bibinfo {author} {\bibfnamefont {K.}~\bibnamefont {Gulyuz}}, \bibinfo {author} {\bibfnamefont {J.}~\bibnamefont {Ariche}}, \bibinfo {author} {\bibfnamefont {G.}~\bibnamefont {Bollen}}, \bibinfo {author} {\bibfnamefont {S.}~\bibnamefont {Bustabad}}, \bibinfo {author} {\bibfnamefont {M.}~\bibnamefont {Eibach}}, \bibinfo {author} {\bibfnamefont {C.}~\bibnamefont {Izzo}}, \bibinfo {author} {\bibfnamefont {S.~J.}\ \bibnamefont {Novario}}, \bibinfo {author} {\bibfnamefont {M.}~\bibnamefont {Redshaw}}, \bibinfo {author} {\bibfnamefont {R.}~\bibnamefont {Ringle}}, \bibinfo {author} {\bibfnamefont {R.}~\bibnamefont {Sandler}}, \bibinfo {author} {\bibfnamefont {S.}~\bibnamefont {Schwarz}},\ and\ \bibinfo {author} {\bibfnamefont {A.~A.}\ \bibnamefont {Valverde}},\ }\bibfield  {title} {\bibinfo {title} {{Determination of the direct double-$\beta$-decay Q value of $^{96}$Zr and atomic masses of $^{90-92, 94, 96}$Zr and $^{92, 94-98, 100}$Mo}},\ }\href {https://doi.org/10.1103/PhysRevC.91.055501}
  {\bibfield  {journal} {\bibinfo  {journal} {Phys. Rev. C}\ }\textbf {\bibinfo {volume} {91}},\ \bibinfo {pages} {055501} (\bibinfo {year} {2015})}\BibitemShut {NoStop}%
\bibitem [{\citenamefont {Ringle}\ \emph {et~al.}(2007{\natexlab{b}})\citenamefont {Ringle}, \citenamefont {Sun}, \citenamefont {Bollen}, \citenamefont {Davies}, \citenamefont {Facina}, \citenamefont {Huikari}, \citenamefont {Kwan}, \citenamefont {Morrissey}, \citenamefont {Prinke}, \citenamefont {Savory}, \citenamefont {Schury}, \citenamefont {Schwarz},\ and\ \citenamefont {Sumithrarachchi}}]{MagneticFieldShift}%
  \BibitemOpen
  \bibfield  {author} {\bibinfo {author} {\bibfnamefont {R.}~\bibnamefont {Ringle}}, \bibinfo {author} {\bibfnamefont {T.}~\bibnamefont {Sun}}, \bibinfo {author} {\bibfnamefont {G.}~\bibnamefont {Bollen}}, \bibinfo {author} {\bibfnamefont {D.}~\bibnamefont {Davies}}, \bibinfo {author} {\bibfnamefont {M.}~\bibnamefont {Facina}}, \bibinfo {author} {\bibfnamefont {J.}~\bibnamefont {Huikari}}, \bibinfo {author} {\bibfnamefont {E.}~\bibnamefont {Kwan}}, \bibinfo {author} {\bibfnamefont {D.~J.}\ \bibnamefont {Morrissey}}, \bibinfo {author} {\bibfnamefont {A.}~\bibnamefont {Prinke}}, \bibinfo {author} {\bibfnamefont {J.}~\bibnamefont {Savory}}, \bibinfo {author} {\bibfnamefont {P.}~\bibnamefont {Schury}}, \bibinfo {author} {\bibfnamefont {S.}~\bibnamefont {Schwarz}},\ and\ \bibinfo {author} {\bibfnamefont {C.~S.}\ \bibnamefont {Sumithrarachchi}},\ }\bibfield  {title} {\bibinfo {title} {{High-precision Penning trap mass measurements of $^{37,38}\mathrm{Ca}$ and their contributions to conserved vector current and
  isobaric mass multiplet equation}},\ }\href {https://doi.org/10.1103/PhysRevC.75.055503} {\bibfield  {journal} {\bibinfo  {journal} {Phys. Rev. C}\ }\textbf {\bibinfo {volume} {75}},\ \bibinfo {pages} {055503} (\bibinfo {year} {2007}{\natexlab{b}})}\BibitemShut {NoStop}%
\bibitem [{\citenamefont {Ehrman}(1951)}]{PhysRev.81.412}%
  \BibitemOpen
  \bibfield  {author} {\bibinfo {author} {\bibfnamefont {J.~B.}\ \bibnamefont {Ehrman}},\ }\bibfield  {title} {\bibinfo {title} {{On the Displacement of Corresponding Energy Levels of ${\mathrm{C}}^{13}$ and ${\mathrm{N}}^{13}$}},\ }\href {https://doi.org/10.1103/PhysRev.81.412} {\bibfield  {journal} {\bibinfo  {journal} {Phys. Rev.}\ }\textbf {\bibinfo {volume} {81}},\ \bibinfo {pages} {412} (\bibinfo {year} {1951})}\BibitemShut {NoStop}%
\bibitem [{\citenamefont {Thomas}(1952)}]{PhysRev.88.1109}%
  \BibitemOpen
  \bibfield  {author} {\bibinfo {author} {\bibfnamefont {R.~G.}\ \bibnamefont {Thomas}},\ }\bibfield  {title} {\bibinfo {title} {{An Analysis of the Energy Levels of the Mirror Nuclei, ${\mathrm{C}}^{13}$ and ${\mathrm{N}}^{13}$}},\ }\href {https://doi.org/10.1103/PhysRev.88.1109} {\bibfield  {journal} {\bibinfo  {journal} {Phys. Rev.}\ }\textbf {\bibinfo {volume} {88}},\ \bibinfo {pages} {1109} (\bibinfo {year} {1952})}\BibitemShut {NoStop}%
\bibitem [{\citenamefont {Firestone}(2009)}]{FIRESTONE20091691}%
  \BibitemOpen
  \bibfield  {author} {\bibinfo {author} {\bibfnamefont {R.}~\bibnamefont {Firestone}},\ }\bibfield  {title} {\bibinfo {title} {{Nuclear Data Sheets for A = 25}},\ }\href {https://doi.org/https://doi.org/10.1016/j.nds.2009.06.001} {\bibfield  {journal} {\bibinfo  {journal} {Nuclear Data Sheets}\ }\textbf {\bibinfo {volume} {110}},\ \bibinfo {pages} {1691} (\bibinfo {year} {2009})}\BibitemShut {NoStop}%
\bibitem [{\citenamefont {Xing}\ \emph {et~al.}(2025)\citenamefont {Xing}, \citenamefont {Luo}, \citenamefont {Zhang}, \citenamefont {Wang}, \citenamefont {Zhou}, \citenamefont {Li}, \citenamefont {Li}, \citenamefont {Yuan}, \citenamefont {Niu}, \citenamefont {Guo}, \citenamefont {Pei}, \citenamefont {Xu}, \citenamefont {de~Angelis}, \citenamefont {Litvinov}, \citenamefont {Blaum}, \citenamefont {Tanihata}, \citenamefont {Yamaguchi}, \citenamefont {Yu}, \citenamefont {Zhou}, \citenamefont {Xu}, \citenamefont {Chen}, \citenamefont {Chen}, \citenamefont {Deng}, \citenamefont {Fu}, \citenamefont {Ge}, \citenamefont {Huang}, \citenamefont {Jiao}, \citenamefont {Li}, \citenamefont {Liao}, \citenamefont {Shi}, \citenamefont {Si}, \citenamefont {Sun}, \citenamefont {Shuai}, \citenamefont {Tu}, \citenamefont {Wang}, \citenamefont {Xu}, \citenamefont {Yan}, \citenamefont {Yuan},\ and\ \citenamefont {Zhang}}]{xing2025z14magicityrevealedmass}%
  \BibitemOpen
  \bibfield  {author} {\bibinfo {author} {\bibfnamefont {Y.~M.}\ \bibnamefont {Xing}}, \bibinfo {author} {\bibfnamefont {Y.~F.}\ \bibnamefont {Luo}}, \bibinfo {author} {\bibfnamefont {Y.~H.}\ \bibnamefont {Zhang}}, \bibinfo {author} {\bibfnamefont {M.}~\bibnamefont {Wang}}, \bibinfo {author} {\bibfnamefont {X.~H.}\ \bibnamefont {Zhou}}, \bibinfo {author} {\bibfnamefont {J.~G.}\ \bibnamefont {Li}}, \bibinfo {author} {\bibfnamefont {K.~H.}\ \bibnamefont {Li}}, \bibinfo {author} {\bibfnamefont {Q.}~\bibnamefont {Yuan}}, \bibinfo {author} {\bibfnamefont {Y.~F.}\ \bibnamefont {Niu}}, \bibinfo {author} {\bibfnamefont {J.~Y.}\ \bibnamefont {Guo}}, \bibinfo {author} {\bibfnamefont {J.~C.}\ \bibnamefont {Pei}}, \bibinfo {author} {\bibfnamefont {F.~R.}\ \bibnamefont {Xu}}, \bibinfo {author} {\bibfnamefont {G.}~\bibnamefont {de~Angelis}}, \bibinfo {author} {\bibfnamefont {Y.~A.}\ \bibnamefont {Litvinov}}, \bibinfo {author} {\bibfnamefont {K.}~\bibnamefont {Blaum}}, \bibinfo {author} {\bibfnamefont {I.}~\bibnamefont
  {Tanihata}}, \bibinfo {author} {\bibfnamefont {T.}~\bibnamefont {Yamaguchi}}, \bibinfo {author} {\bibfnamefont {Y.}~\bibnamefont {Yu}}, \bibinfo {author} {\bibfnamefont {X.}~\bibnamefont {Zhou}}, \bibinfo {author} {\bibfnamefont {H.~S.}\ \bibnamefont {Xu}}, \bibinfo {author} {\bibfnamefont {Z.~Y.}\ \bibnamefont {Chen}}, \bibinfo {author} {\bibfnamefont {R.~J.}\ \bibnamefont {Chen}}, \bibinfo {author} {\bibfnamefont {H.~Y.}\ \bibnamefont {Deng}}, \bibinfo {author} {\bibfnamefont {C.~Y.}\ \bibnamefont {Fu}}, \bibinfo {author} {\bibfnamefont {W.~W.}\ \bibnamefont {Ge}}, \bibinfo {author} {\bibfnamefont {W.~J.}\ \bibnamefont {Huang}}, \bibinfo {author} {\bibfnamefont {H.~Y.}\ \bibnamefont {Jiao}}, \bibinfo {author} {\bibfnamefont {H.~F.}\ \bibnamefont {Li}}, \bibinfo {author} {\bibfnamefont {T.}~\bibnamefont {Liao}}, \bibinfo {author} {\bibfnamefont {J.~Y.}\ \bibnamefont {Shi}}, \bibinfo {author} {\bibfnamefont {M.}~\bibnamefont {Si}}, \bibinfo {author} {\bibfnamefont {M.~Z.}\ \bibnamefont {Sun}}, \bibinfo
  {author} {\bibfnamefont {P.}~\bibnamefont {Shuai}}, \bibinfo {author} {\bibfnamefont {X.~L.}\ \bibnamefont {Tu}}, \bibinfo {author} {\bibfnamefont {Q.}~\bibnamefont {Wang}}, \bibinfo {author} {\bibfnamefont {X.}~\bibnamefont {Xu}}, \bibinfo {author} {\bibfnamefont {X.~L.}\ \bibnamefont {Yan}}, \bibinfo {author} {\bibfnamefont {Y.~J.}\ \bibnamefont {Yuan}},\ and\ \bibinfo {author} {\bibfnamefont {M.}~\bibnamefont {Zhang}},\ }\href {https://arxiv.org/abs/2503.01380} {\bibinfo {title} {{$Z=14$ Magicity Revealed by the Mass of the Proton Dripline Nucleus $^{22}$Si}}} (\bibinfo {year} {2025}),\ \Eprint {https://arxiv.org/abs/2503.01380} {arXiv:2503.01380 [nucl-ex]} \BibitemShut {NoStop}%
\end{thebibliography}%
\vspace{1mm}

\end{document}